\documentclass{article}

\usepackage{arxiv}

\usepackage[utf8]{inputenc} % allow utf-8 input
\usepackage[T1]{fontenc}    % use 8-bit T1 fonts
\usepackage{hyperref}       % hyperlinks
\usepackage{url}            % simple URL typesetting
\usepackage{booktabs}       % professional-quality tables
\usepackage{amsfonts}       % blackboard math symbols
\usepackage{nicefrac}       % compact symbols for 1/2, etc.
\usepackage{microtype}      % microtypography
\usepackage{lipsum}
\usepackage{graphicx}
\usepackage{subcaption}
\usepackage{float}
\usepackage{xcolor}
\title{The VIX index under scrutiny of machine learning techniques and neural networks}

\author{
  Ali Hirsa \\
  IEOR Department\\
  Data Science Institute \\
  Columbia University\\
  \texttt{ali.hirsa@columbia.edu} \\
   \And
 Joerg Osterrieder\footnotemark[\value{footnote}] \\
  School of Engineering\\
  Zurich University of Applied Sciences\\
  Winterthur, Switzerland \\
  \texttt{joerg.osterrieder@zhaw.ch} \\
   \And
 Branka Hadji Misheva \\
  School of Engineering\\
  Zurich University of Applied Sciences\\
  Winterthur, Switzerland \\
  \texttt{branka.hadjimisheva@zhaw.ch} \\
   \And 
 Wenxin Cao \\
  IEOR Department\\
  Columbia University\\
  New York, NY 10027 \\
  \texttt{wc2695@columbia.edu} \\
   \And  
  Yiwen Fu \\
  IEOR Department \\
  Columbia University\\
  New York, NY 10027 \\
  \texttt{yf2507@columbia.edu} \\
   \And
   Hanze Sun \\
   IEOR Department\\
  Columbia University\\
  New York, NY 10027 \\
  \texttt{hs3237@columbia.edu}
   \And
 Kin Wai Wong \\
  IEOR Department\\
  Columbia University\\
  New York, NY 10027 \\
  \texttt{kw2874@columbia.edu} \\
  }

\begin{document}
\maketitle
\footnotetext{Financial support by the Swiss National Science Foundation within the project “Mathematics and Fintech - the next revolution in the digital transformation of the Finance industry” is gratefully acknowledged by the corresponding author. 
This research has also received funding from the European Union's Horizon 2020 research and innovation program FIN-TECH: A Financial supervision and Technology compliance training programme under the grant agreement No 825215 (Topic: ICT-35-2018, Type of action: CSA).
Furthermore, this article is based upon work from COST Action 19130 Fintech and Artificial Intelligence in Finance, supported by COST (European Cooperation in Science and Technology), www.cost.eu (Action Chair: Joerg Osterrieder).\newline
The authors are grateful to Stephan Sturm, management committee members of the COST (Cooperation in Science and Technology) Action Fintech and Artificial Intelligence in Finance as well as speakers and participants of the 5th European COST Conference on Artificial Intelligence in Finance and Industry, which took place at Zurich University of Applied Sciences, Switzerland, in September 2020.}

\begin{abstract}
The CBOE Volatility Index, known by its ticker symbol VIX, is a popular measure of the market’s expected volatility on the S\&P 500 Index, calculated and published by the Chicago Board Options Exchange (CBOE). It is also often referred to as the fear index or the fear gauge. The current VIX index value quotes the expected annualized change in the S\&P 500 index over the following 30 days, based on options-based theory and current options-market data. Despite its theoretical foundation in option price theory, CBOE’s Volatility Index is prone to inadvertent and deliberate errors because it is weighted average of out-of-the-money calls and puts which could be illiquid. Many claims of market manipulation have been brought up against VIX in recent years. This paper discusses several approaches to replicate the VIX index as well as VIX futures by using a subset of relevant options as well as neural networks that are trained to automatically learn the underlying formula. Using subset selection approaches on top of the original CBOE methodology, as well as building machine learning and neural network models including Random Forests, Support Vector Machines, feed-forward neural networks, and long short-term memory (LSTM) models, we will show that a small number of options is sufficient to replicate the VIX index. Once we are able to actually replicate the VIX using a small number of S\&P options we will be able to exploit potential arbitrage opportunities between the VIX index and its underlying derivatives. The results are supposed to help investors to better understand the options market, and more importantly, to give guidance to the US regulators and CBOE that have been investigating those manipulation claims for several years.

\end{abstract}

% keywords can be removed

\section{Introduction}
The VIX is based on S\&P 500 option quotes every 15 seconds and is intended to provide an indication of the fair market price of the expected volatility of the S\&P 500 index over the next 30 days \cite{CBOE}. Although it is not directly tradeable, there are many derivatives on the index, including both options and futures. For many years, the VIX has been questioned since it is vulnerable to liquidity and price fluctuations in the underlying options market as well as the related derivatives market \cite{Osterrieder2019}. Thus, through replicating the VIX index using approaches different that the official CBOE methodology, we want to investigate whether it can be represented and influenced by a small subset of options. Compared to selecting hundreds of options by the official methodology, we analyze the replication performance by using a fraction of it. If there exists a way to replicate the index, with a high accuracy, using a subset of options, this might imply that the VIX is impacted by a small number of options and it might be possible to exploit potential arbitrage opportunities between the VIX and its derivatives.

To begin with, we perform the replication on both a daily and an intraday scale based on the official methodology and formulas outlined in the VIX White Paper (for more details see \cite{CBOE}). This represents our benchmark model. Next, we replicate the index with a subsets of options. After analysing the results, we select the subset with the best performance as the input of our neural network models, including feed-forward neural networks and LSTM models. Our findings show that the VIX index can be replicated using a subset of options and a neural network which learns the dependencies between the input and output features, i.e. learns the official formula. Furthermore, we outline how one can exploit potential arbitrage opportunities without having to trade hundreds of S\&P 500 options (\cite{CBOE}).

The remainder of the paper is organized as follows: Section 2 provides a literature review of relevant studies that discuss different perspective on the claim of VIX manipulation. Section 3 describes the data and methodology we use to replicate VIX, as well as the subset selection approaches to show the significance and influence of a number of options. In Section 4, we describe our design and implementation of a basic neural network and LSTM models, as well as the replication results. Finally, Section 5 discusses the conclusion of our findings and potential future extensions.

\section{Literature Review }
The VIX and its derivatives have been extensively discussed in the literature. In Osterrieder et al. (2019) \cite{Osterrieder2019}, the author gives a thorough analysis from both a theoretical and empirical perspective. By deriving the VIX formula and empirically examining the individual components of the formula, the authors conclude that the current formula is not accurate enough and thus opens the door for potential manipulators. Indeed, futures manipulation has long been studied in the literature. Kumar and Seppi (1992) \cite{Kumar1992} investigate the susceptibility of futures markets to price manipulation in a two-period model with asymmetric information and cash-settled futures contracts. They interpret manipulation as a form of endogenous noise trading which can arise in multi-period security markets. In Pimbley (2018) \cite{Pimbley2018}, the authors examines the VIX formula in detail and give explanations of how the approximations of the CBOE method promote susceptibility to errors and manipulations. The paper also discusses how to remove or limit these approximations and gives a review of numerical results of the improvements. 

On the other side of the discussion, the literature also offers papers which do not support the argument of manipulation of the VIX. In Saha (2018) \cite{Saha2018}, the authors construct a regression model with explanatory variables that are exogenous to the index and find that the movements in the daily levels of the index are explained by market fundamentals instead of manipulation. By examining the futures on its expiration days, the paper demonstrates that the VIX closing values and settlement prices on those days are consistent with normal market forces and not artificial. Also, the analysis conducted by Wei (2019) \cite{analysisgroup} suggests that manipulation techniques are not easily applied in a real-world setting. He finds that the total trading volume needs to be extremely large in order to manipulate the VIX. Furthermore, reasons other than manipulation may explain the trading activity on any particular day. 

In addition, there is also substantial research which aims at studying the VIX and its underlying by using novel techniques. In Osterrieder (2019) \cite{Osterrieder2019}, the authors propose a machine learning approach, specifically an LSTM model with one layer to predict the VIX and shows that ten options are sufficient to achieve a very high accuracy. 

Our paper contributes the literature by:providing a thorough review of the VIX methodology provided by CBOE and further expanding the topic by using machine learning and deep learning techniques (including feature selection, basic neural network models and LSTM models) to replicate VIX as well as VIX futures and studying their characteristics.

\section{Constructing the VIX Index}
In this section, we describe how we replicated VIX using the original methodology and formulas documented in the CBOE VIX White Paper. In addition, the replication of VIX was done with both daily end-of-day (EOD) data and intraday minute-by-minute data, and our results are satisfactory despite some minor discrepancies caused by small differences between our replication and the original methodology. 

\subsection{Data}
The data consists of all the S\&P 500 (SPX) options in the market. There are two sets of data: an EOD SPX options quote dataset from January 2017 to March 2018, and a minute-by-minute SPX options quote dataset from January 2018 to February 2018. The former is used for the daily VIX replication while the latter is used for the intraday replication. The EOD dataset contains the closing price released at 3:15 p.m. Chicago time. The VIX index is available based on 15-second intervals, as per the CBOE methodology.

\subsection{Methodology for replicating the VIX Index}
Our replication methodology follows the procedures and rules documented in the CBOE VIX White Paper. The components of the VIX Index are near-term and next-term SPX put and call options with more than 23 days and less than 37 days to expiration. All the details can be found in the official CBOE VIX white paper (\cite{CBOE}). 
There are two differences between our replication and the official methodology. The first difference is the risk-free interest rate used in the calculation of the forward price. Due to its small influence during the computation, it is set to a constant of 1.3\% in our intraday implementation and 1\% in our daily replication with the purpose of providing an easier comparison of results. Both values are determined as the approximate mean of the risk-free rate for the corresponding time intervals. The second difference is that our implementation does not include the VIX Index Filtering Algorithm. As described in the VIX White Paper, CBOE uses this filtering algorithm to address large potential jumps in the VIX index. Due to its small impact (on average), we are not including this algorithm in our replication.

\subsection{Results \& Validation}
For the daily replication, we compare our replicated values with the true EOD VIX values. Our daily replication method leads to an MSE of $0.0025$, a MAE of $0.0238$, and a mean percentage error of $0.184\%$.

\begin{figure}[h]
\begin{center}
\includegraphics[width=0.8\linewidth, height=7cm]{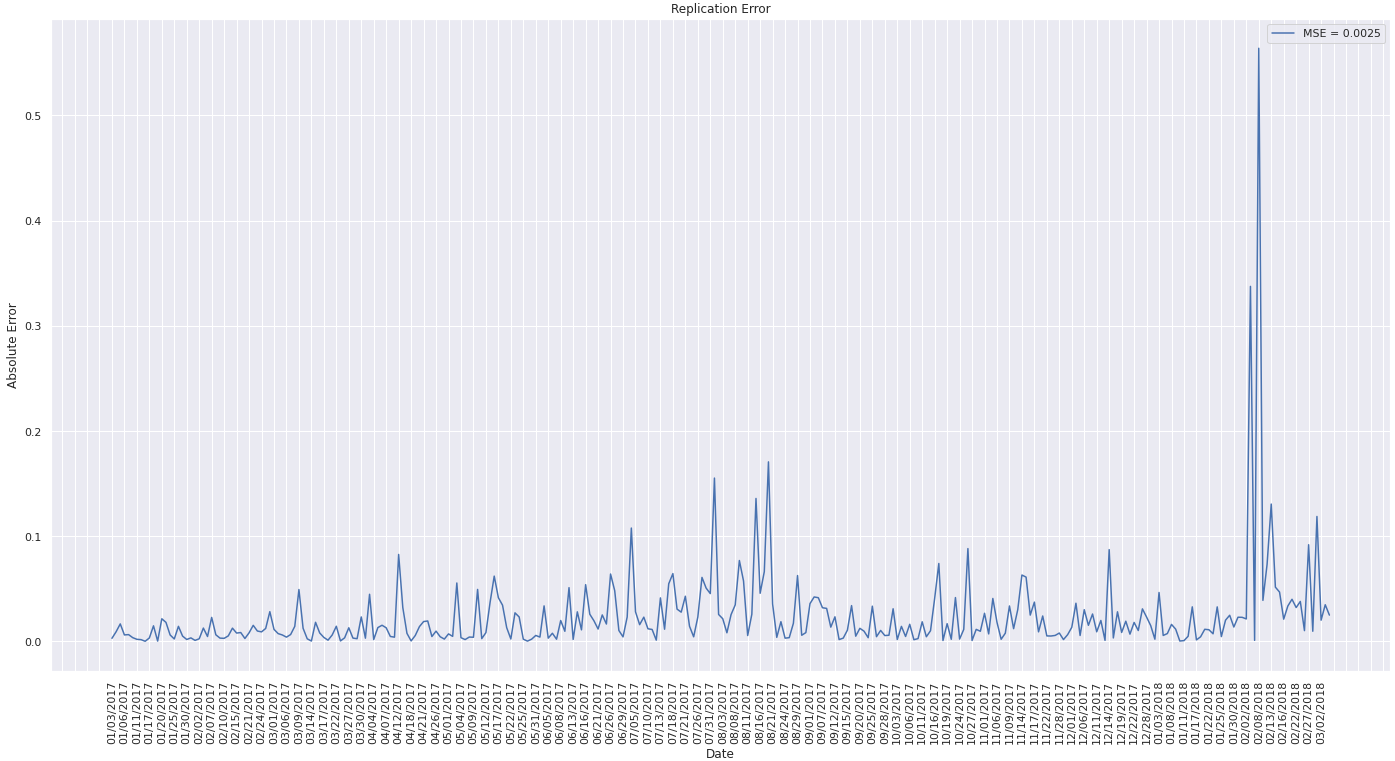}
\caption{Replication error of the VIX index for daily close values}
\label{fig:subim1d}
\end{center}
\end{figure}

\begin{figure}[h]
\begin{center}
\includegraphics[width=0.8\linewidth, height=7cm]{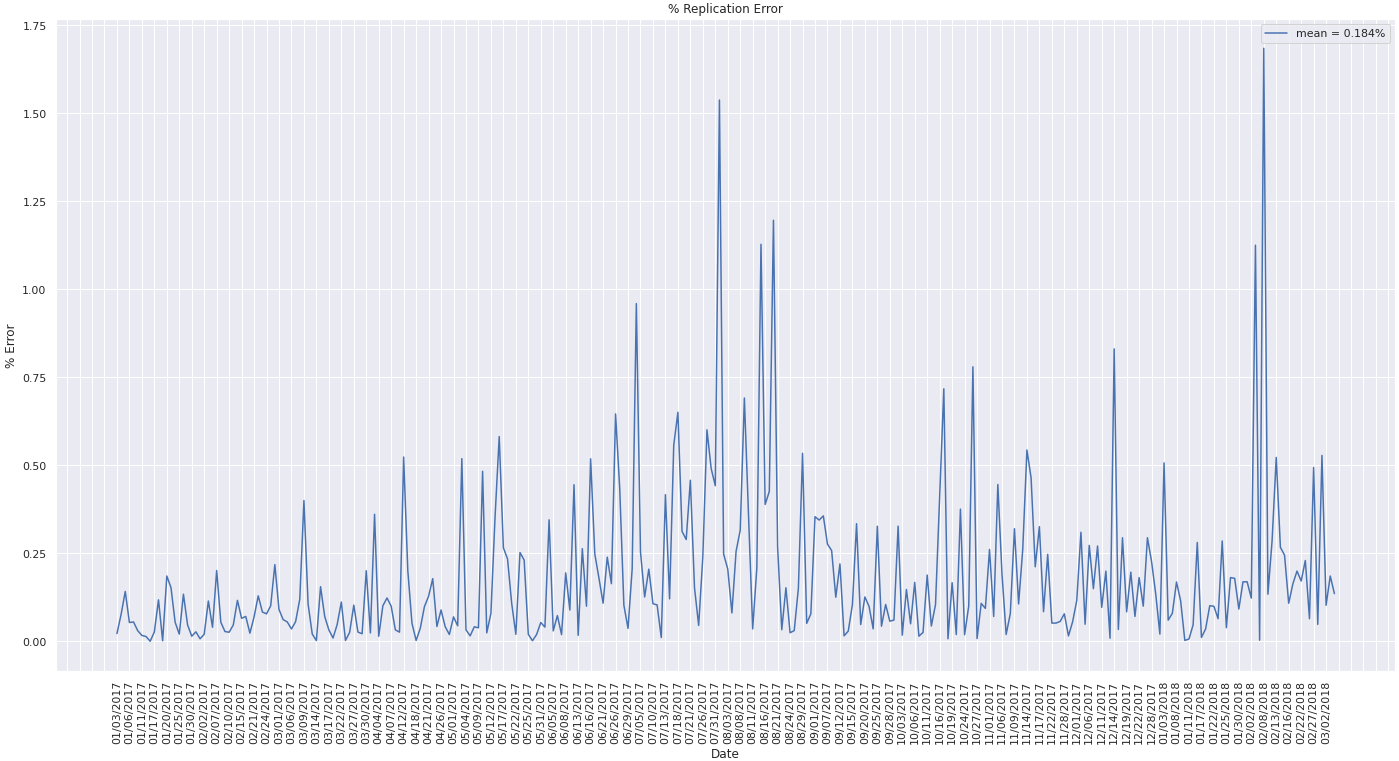}
\caption{Percentage replication error of the VIX for daily close values}
\label{fig:subim1e}
\end{center}
\end{figure}

To better evaluate the original VIX methodology, we also perform an intraday replication. As for the replication performance, we validated our replicated values with the real VIX values and obtain an MSE of $0.0325$, an MAE of $0.0637$, and a mean percentage error of $0.336\%$.

\begin{figure}[h]
\begin{center}
\includegraphics[width=0.8\linewidth, height=7cm]{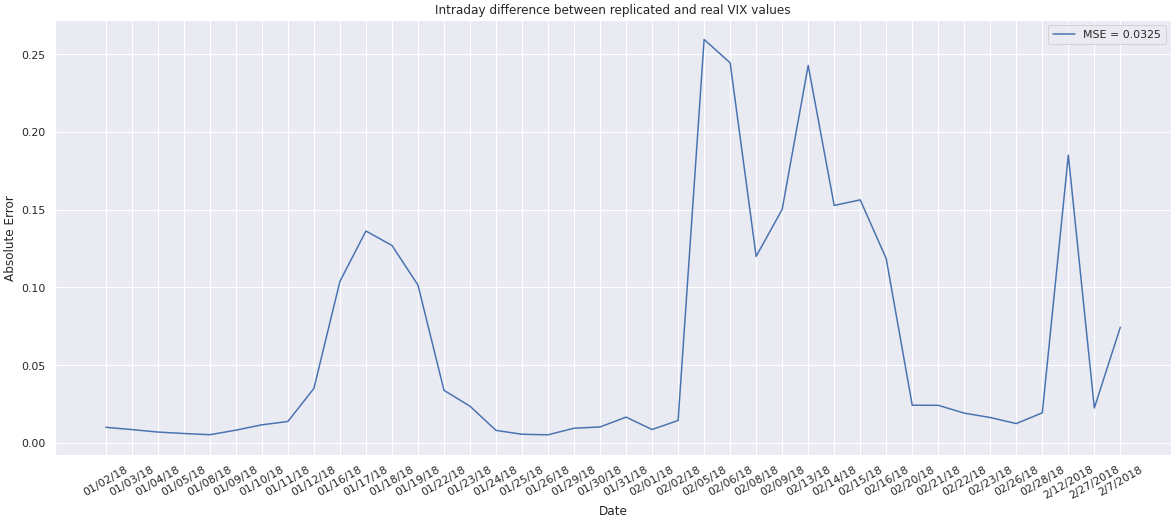}
\caption{Replication error of the VIX index for intraday minute-by-minute values}
\label{fig:subim1f}
\end{center}
\end{figure}
\begin{figure}[h]
\begin{center}
\includegraphics[width=0.8\linewidth, height=7cm]{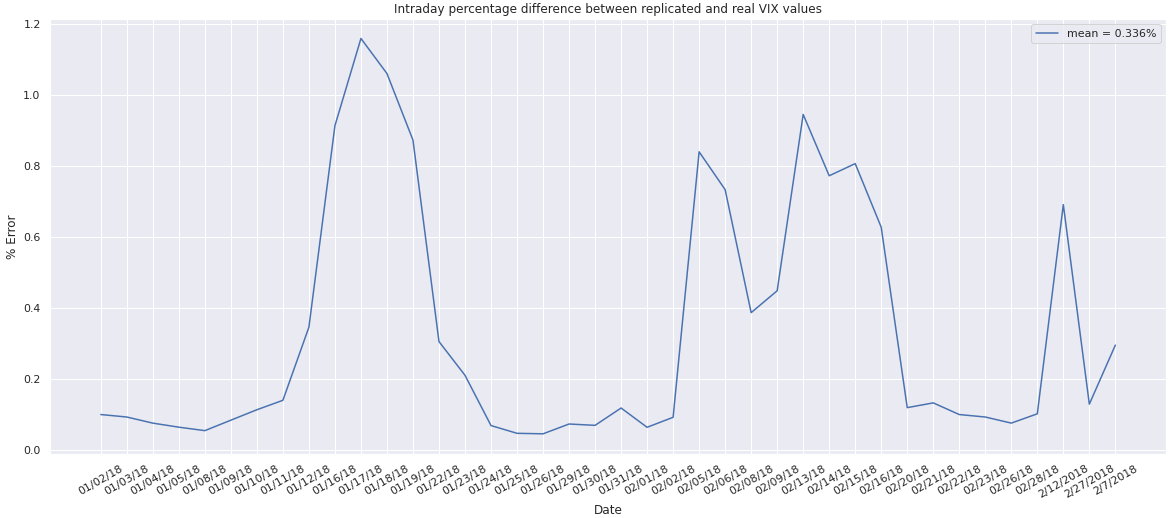}
\caption{Percentage replication error of the VIX index for intraday minute-by-minute values}
\label{fig:subim1g}
\end{center}
\end{figure}

Although the MSE and MAE of both the daily and the intraday replication are acceptable, there are still some minor discrepancies between our replicated values and the actual values. The causes of the discrepancies are the following three. The first one are time-stamp differences. The intraday S\&P 500 options dataset records all the information at every minute starting from 09:31:00 EST. On the other hand, the intraday VIX spot value dataset not only does not record the information starting at 09:31:00 ET, but also at different seconds on every day. For example, the information was recorded starting from 09:31:29 on January 2, 2018, while starting from 09:31:17 EST on January 10, 2018.  Such varying time stamps on every day cause difficulties in adjusting the intraday replication to have matching time stamps. Moreover, on a daily level, the VIX spot price data normally does not have its EOD record at exactly 3:15 p.m. EST, rather it fluctuates between 1 to 15 seconds earlier. In summary, the time-stamp difference between the recording time of the options data and the VIX spot value data explains the discrepancies in both the daily and the intraday replication. The second and third sources of error are the usage of a constant risk-free interest rate and the exclusion of the VIX index filtering algorithm. As mentioned before, we have decided to make those two changes in order to simplify the comparison and understanding of the results, knowing that they will have small influences on the replication performance. 

Besides using the MSE and the MAE to analyse our replications, we have also obtained the correlation between the replicated values and the real VIX values as 0.999 for both replications. The high values indicate that the replicated values are strongly correlated to the real values. In addition, one of the findings is that the number of put options selected during the option selection process is generally higher than the number of call options, an expected results based on the well-known richness of out-of-the-money put options. We have re-computed the number of call and put options selected during the intraday replication minute by minute, and the average number of put options selected for every minute is 258 while the average number of call options is 109. This reflects the financial market dynamics that investors are looking for downside protection with put options, causing more put options to have nonzero bid price and thereby being selected by the VIX methodology. The following graph shows the average number of put and call options selected on every day, calculated by taking the average of all the minute-by-minute counts on that day. 
\begin{figure}[h]
\begin{center}
\includegraphics[width=0.8\linewidth, height=7cm]{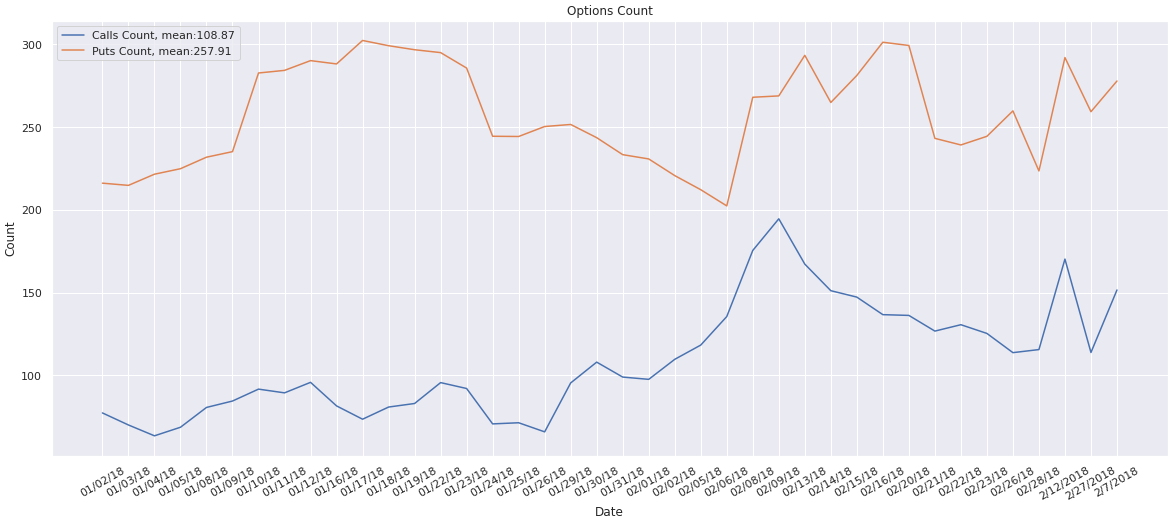}
\caption{Count of call and put options selected in intraday replication}
\label{fig:subim1h}
\end{center}
\end{figure}
\subsection{Selecting the most relevant options}
In this section, we describe how we select a subset of options rather than using all the options selected based on the original methodology to replicate VIX. In other words, we only use a portion of options selected by the official methodology, while simultaneously still using the official formulas for computing the VIX index. There are two approaches we implemented using a subset selection, and we discuss their methodologies and compare their performance. 

\subsubsection{Selecting a subset of options every minute}
This approach is to select a subset of options that are nearest to the at-the-money implied underlying price $K_0$, the original criteria in the VIX White Paper, for both near-term and next-term options in every minute. For implementation, after selecting all the options using the original methodology, we are only keeping a given number of near-term and next-term options that are closest to the near-term and next-term $K_0$ respectively. In our implementation, we have computed the VIX based on $1, 2, 3, 4, 5, 10, 25, 50, 100, 200$, and all available options.

Next, since we are using only a subset of options and the VIX index is a weighted average of the option prices, we have to adjust the corresponding result to match our reduced sample of options. We first scale the corresponding result by a constant value, to have equal means of both time-series, before examining the error. Despite very good replication results, a clear benefit from using more options is still visible.
\begin{figure}[h]
\includegraphics[width=0.3\linewidth, height=5.8cm]{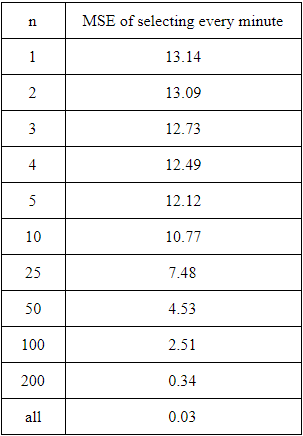}
\centering
\caption{MSE of selecting a subset of options every minute with different numbers of options (n) to replicate the VIX index (after scaling)}
\label{fig:subim1i}
\end{figure}
%
%
% \begin{table}[h]
% \begin{center}
% \begin{tabular}{|c|c|} \hline
% $n$ & MSE of selecting every minute \\ \hline\hline
% 1 & 13.14  \\ \hline
% 2 & 13.09  \\ \hline
% 3 &  12.73 \\ \hline
% 4 &  12.49 \\ \hline
% 5 &  12.12 \\ \hline
% 10 & 10.77 \\ \hline
% 25 &  7.48 \\ \hline
% 50 &  4.53 \\ \hline
% 100 & 2.51 \\ \hline
% 200 & 0.34 \\ \hline
% all & 0.03 \\ \hline
% \end{tabular}
% \end{center}
% \caption{MSE of selecting a subset of options every minute with different numbers of options (n) to replicate the VIX index (after scaling)}
% \label{fig:subim1i}
% \end{table}%
%
%
\begin{figure}[h]
\begin{center}
\includegraphics[width=0.8\linewidth, height=7cm]{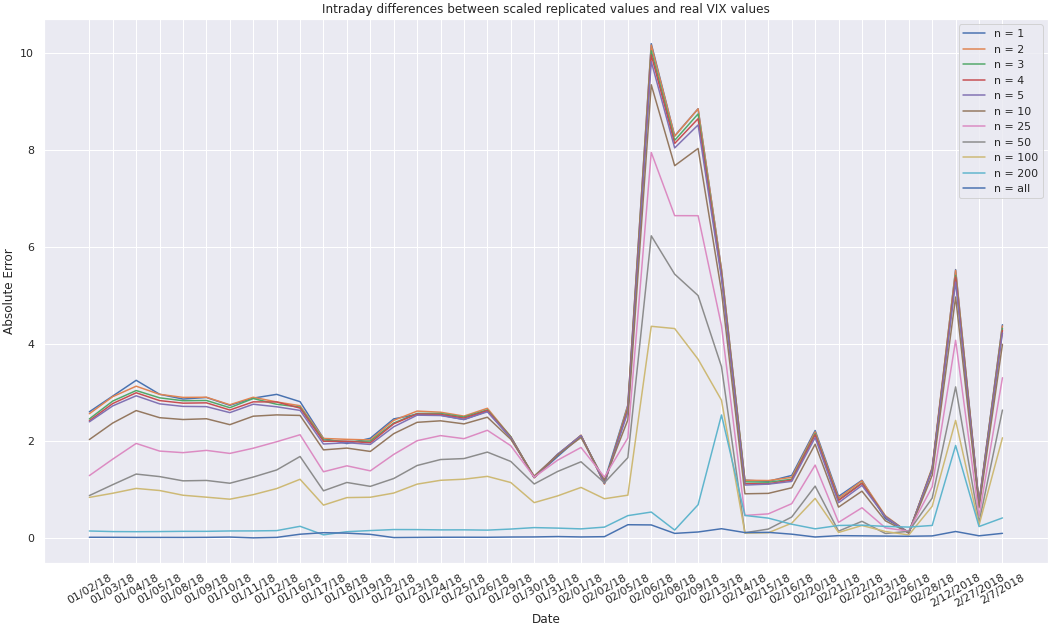}
\caption{Error of selecting a subset of options every minute with different numbers of options (n) to replicate the VIX index (after scaling)}
\label{fig:subim1j}
\end{center}
\end{figure}
We also compute the correlation between the replicated values using a subsets of options and the real VIX values. Once again the high correlation proves that a small subset of options will give a good performance in replicating the VIX index.
\begin{figure}[h]
\includegraphics[width=0.30\linewidth, height=6cm]{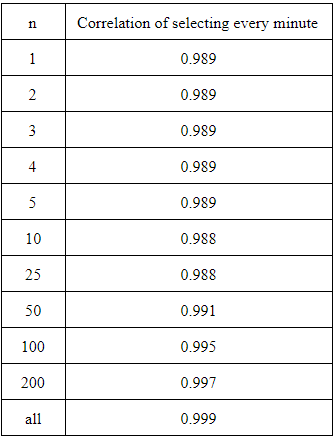}
\centering
\caption{Correlation of selecting a subset of options every minute with different numbers of options (n) to replicate the VIX index}
\label{fig:subim1l}
\end{figure}
\subsubsection{Selecting a subset of options at the beginning of every day}
We proceed as in the previous section, with the exception of fixing the subset of options at the beginning of the day at 09:31:00 EST. As expected, replication results are not as good, but still sufficient for actual trading.
\begin{figure}[h]
\includegraphics[width=0.3\linewidth, height=5.5cm]{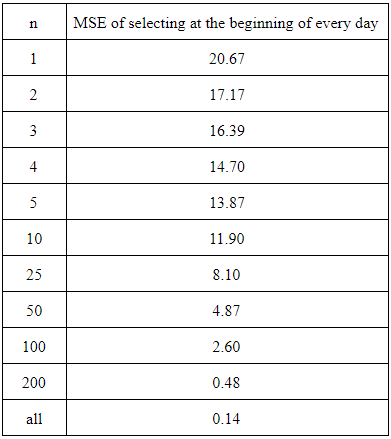}
\centering
\caption{MSE of selecting a subset of options at the beginning of every day with different numbers of options (n) to replicate the VIX index (after scaling)}
\label{fig:subim1m}
\end{figure}
\begin{figure}[h]
\begin{center}
\includegraphics[width=0.8\linewidth, height=7cm]{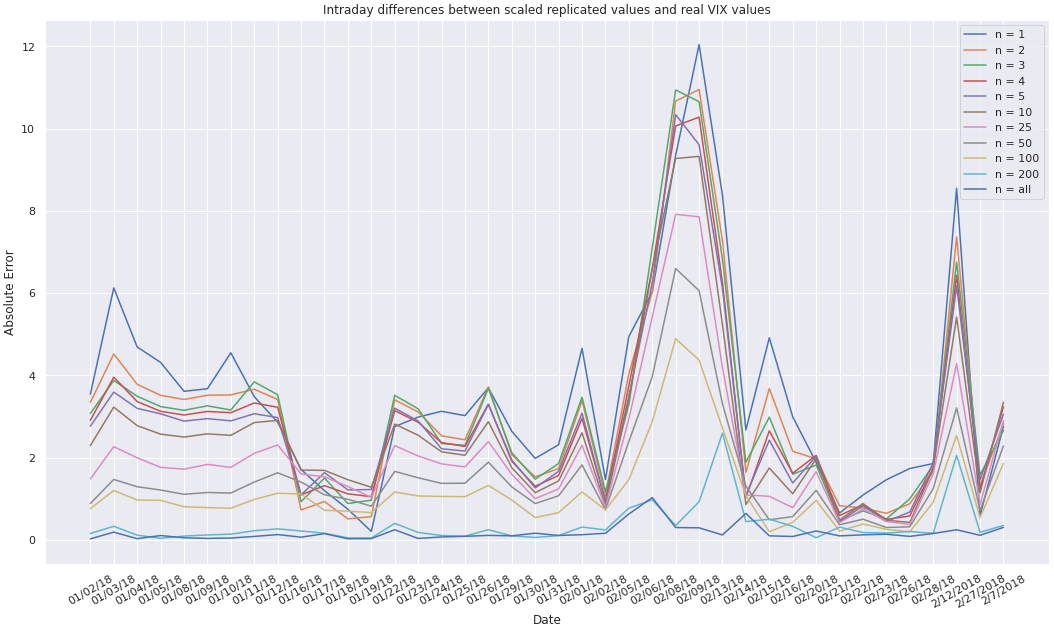}
\caption{Error of selecting a subset of options at the beginning of every day with different numbers of options (n) to replicate the VIX (after scaling)}
\label{fig:subim1n}
\end{center}
\end{figure}
As before, we compute the correlation between the replicated values and the real VIX values.
%
%
% \begin{table}[h]
% \begin{center}
% \begin{tabular}{|c|c|} \hline
% $n$ & Correlation of selecting every minute \\ \hline\hline
% 1 & 0.989  \\ \hline
% 2 & 0.989  \\ \hline
% 3 & 0.989 \\ \hline
% 4 & 0.989 \\ \hline
% 5 & 0.989 \\ \hline
% 10 & 0.988 \\ \hline
% 25 & 0.988\\ \hline
% 50 & 0.991 \\ \hline
% 100 & 0.995 \\ \hline
% 200 & 0.997 \\ \hline
% all & 0999 \\ \hline
% \end{tabular}
% \end{center}
% \caption{Correlation of selecting a subset of options at the beginning of every day with different numbers of options(n) to replicate VIX}
% \label{fig:subim1o}
% \end{table}%
%
%
\begin{figure}[h]
\includegraphics[width=0.32\linewidth, height=6cm]{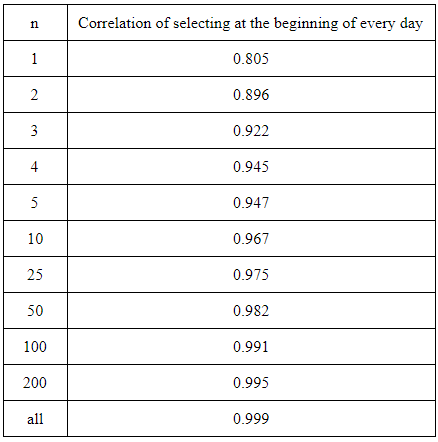}
\centering
\caption{Correlation of selecting a subset of options at the beginning of every day with different numbers of options(n) to replicate VIX}
\label{fig:subim1o}
\end{figure}
Comparing both approaches, we conclude that selecting a new subset of options every minute gives a better replication performance, leading to a lower MSE. As a result, we are using this approach to generate the input of the neural network models in the following section.
\begin{figure}[h]
\includegraphics[width=0.45\linewidth, height=6cm]{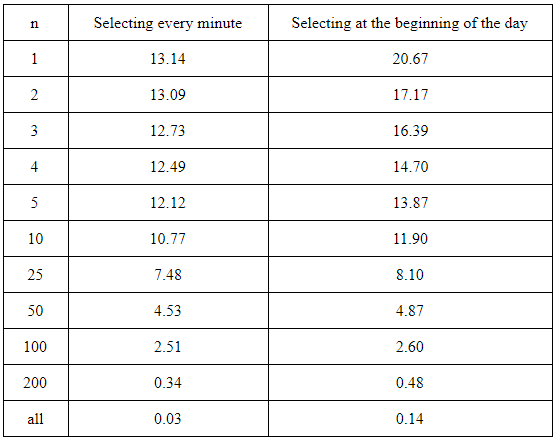}
\centering
\caption{Comparison of the replication MSE of both approaches}
\label{fig:subim1p}
\end{figure}
\section{Neural networks for the VIX and VIX futures}
To better understand the VIX index, we use basic neural network models and LSTM models to replicate the intraday VIX index. 

\subsection{A basic neural network}
Before applying the more sophisticated model, we first build a basic neural network model and study its performance and the relation between the VIX index and its underlying options.
\subsubsection{Data}\label{sec:data}
The data used in the model is the intraday S\&P500 data and VIX index from January 2, 2018 to February 28, 2018. Our subset of options is based on the selection as per the CBOE methodology. For every minute, we choose 52 options, including both near-term and next term, as features, in particular every third option based on strike price. 

\subsubsection {Methodology and Result}

We first build a basic neural network model using the option prices as the only feature (Model 1), so the inputs of the model are the subset of option prices at every minute every day in the two months period and the per minute VIX index in the same period. 
The model contains two dense layers with each having 64 nodes, and one output layer (see Table 1). The activation function is the rectified linear activation function, the loss function is mean squared error and the optimizer is Adam. We randomly assign 70\% of the data as the training data and the remaining 30\% as the test data.
After training the model and doing the prediction, we got the in-sample MSE of $0.064$ and out-of-sample MSE of $0.095$. We also performed 5-fold cross- validation and obtained an MSE of $0.162$.

In addition, we set up additional regressors including a decision tree regressor, random forest regressor and a support vector regressor as benchmarks. Both the decision tree regressor and the random forest regressor were implemented with a maximum number of leaf nodes of eight and the random forest regressor with ten trees. For the support vector regressor, we used a radial basis function kernel. The MSE of the decision tree regressor, random forest regressor, and support vector regressor, are $1.188, 1.080,$ and $0.337$ respectively. Compared with these regressors, the neural network model has the smallest MSE.

Besides option prices, we also trained another neural network model (Model 2) and we used the option prices plus the option greeks including theta and vega as the features. Since the number of the features increases, we built a slightly more complex model than before to fit the data. We expanded the layers to 4 and the number of nodes to 128 and as a result we obtained an in-sample MSE of 0.062 and an out-of-sample MSE of 0.090. The 5-fold cross-validation gives a MSE of 0.126. Thus, using the additional features as inputs, the neural network model has an smaller MSE than all of the previously tested approaches. We also set up additional regressors as benchmarks. The MSE of the decision tree regressor, random forest regressor, and support vector regressor are 1.051, 0.925 and 0.371, respectively. The MSE of the neural network model with more features is still the smallest one compared to the benchmarks.
\begin{table}
\centering
\begin{subtable}{0.5\textwidth}
\centering
 \caption{Basic neural network,  Model 1}
  \centering
  \begin{tabular}{lll}
    \toprule
    \cmidrule(r){1-2}
    Layer (type)  & Output Shape & Param \# \\
    \midrule
   flatten (Flatten)& (None, 52) &0       \\
    dense\_1 (Dense) &  (None, 64) & 3392 \\
   dense\_2 (Dense)  &   (None, 64) & 4160  \\
   dense\_3 (Dense) &(None, 1)  &  65 \\
    \midrule
    Total params: 7,617\\
    Trainable params: 7,617\\
    Non-trainable params: 0\\
    \bottomrule
  \end{tabular}
  \label{tab:table}
\end{subtable}%

\begin{subtable}{0.5\textwidth}
\centering
 \caption{Basic neural network, Model 2}
  \centering
  \begin{tabular}{lll}
    \toprule
    \cmidrule(r){1-2}
    Layer (type)  & Output Shape & Param \# \\
    \midrule
   flatten (Flatten)& (None, 156) &0       \\
    dense\_1 (Dense) &  (None, 128) & 20096 \\
   dense\_2 (Dense)  &   (None, 128) & 16512  \\
   dense\_3 (Dense) &  (None, 128) & 16512 \\
   dense\_4 (Dense)  &   (None, 128) & 16512  \\
   dense\_5 (Dense) &(None, 1)  &  129 \\
    \midrule
    Total params: 69,761\\
    Trainable params: 69,761\\
    Non-trainable params: 0\\
    \bottomrule
  \end{tabular}
  \label{tab:table}
\end{subtable}

\end{table}

\begin{figure}[h]
\begin{center}
\includegraphics[width=0.5\linewidth, height=5cm]{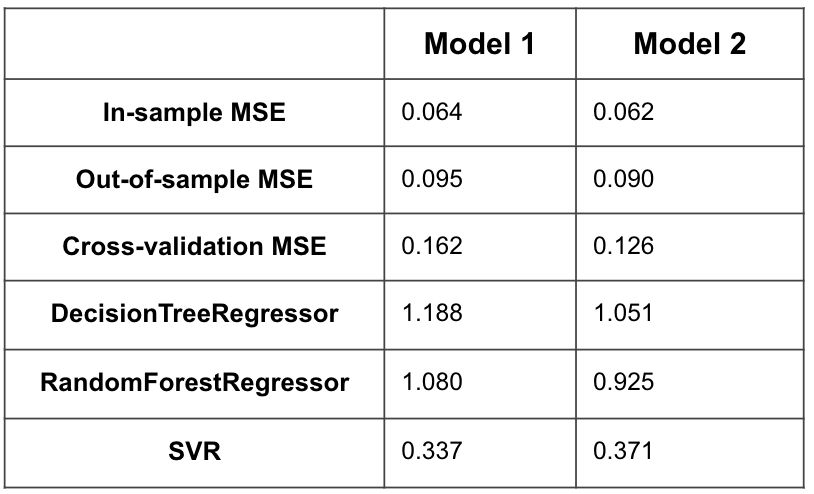}
\caption{Summary: Performance of both models}
\label{fig:subim1q}
\end{center}
\end{figure}

\begin{figure}[h]
\begin{subfigure}{0.5\textwidth}
\includegraphics[width=0.9\linewidth, height=5cm]{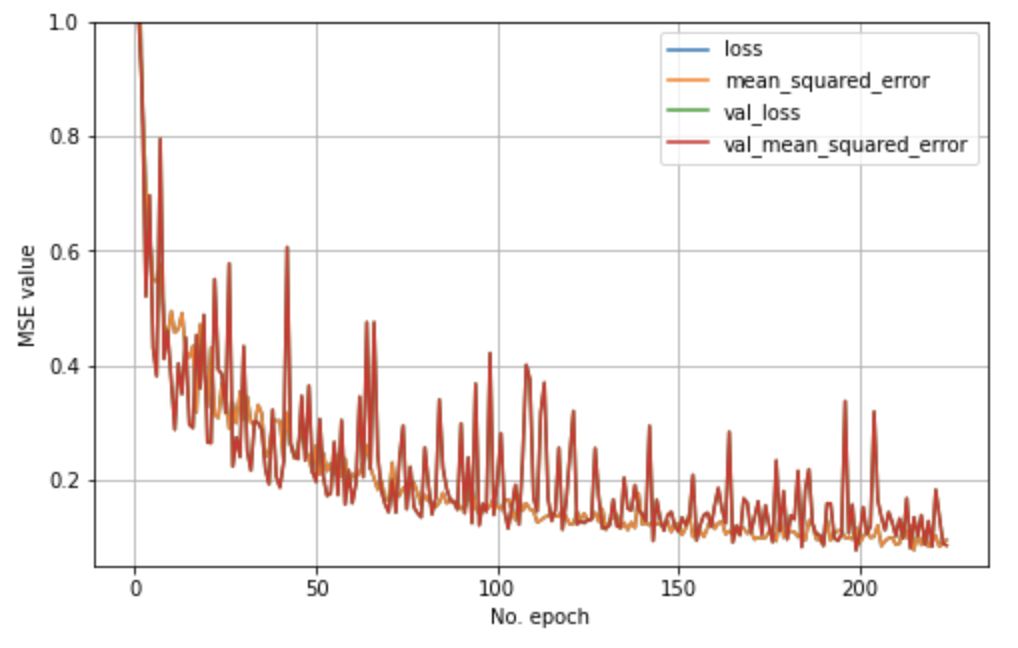}
\caption{Training Loss for the basic neural network, model 1}
\label{fig:subim1r}
\end{subfigure}
\begin{subfigure}{0.5\textwidth}
\includegraphics[width=0.9\linewidth, height=5cm]{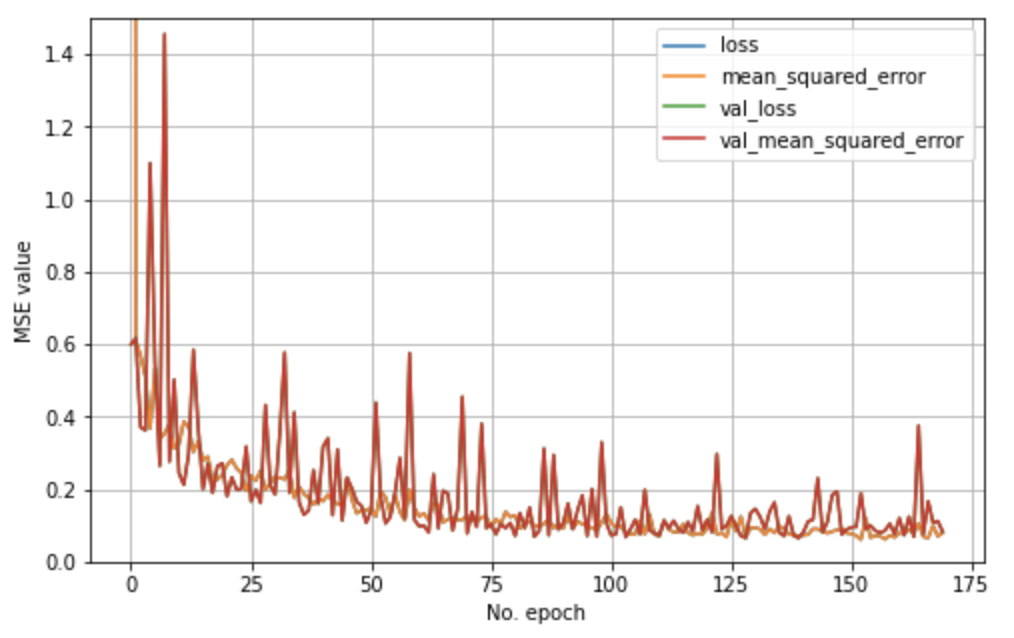}
\caption{Training Loss for the basic neural network, model 2}
\label{fig:subim15}
\end{subfigure}
\end{figure}

\subsection{LSTM}
Lastly, we use Long Short Term Memory networks (LSTM) to study the long-term dependencies of S\&P options data.

LSTM is a type of a recurrent neural network (RNN), a neural network that persists information of previous events to inform later ones. However, one of the drawbacks of RNNs is that they are unable to learn to connect the information when there is a larger context to process. An LSTM is a specialized RNN that is capable of learning long term dependencies.

\subsubsection {Data and Methodology}
The data used in the model is the same as described in Section \ref{sec:data} We use the intraday option price data as our input, and the VIX index value as the value to predict. After splitting the data into training and test set, we scaled the data to the unit interval. 

We analysed two different model architectures, a flat model and a multi-layer model. For each model architecture, we also investigated two different input data sets. One of them selects options by order, and the other selects every third option as input. The latter ensures our input to be spread more evenly across the price to eliminate duplicate information.

For each model, we used a learning rate of $0.000005$ and $300$ epochs. We also introduced two callbacks, an early stopping callback with a patience of 1500 on the validation loss and a tensorboard callback for better visualizations.

\subsubsection {Model 1: The Flat Model} 
The flat model contains an LSTM layer with 1000 output units and a dense layer with one neuron (see Table 2). 
  
\begin{table}
 \caption{LSTM Model: Flat}
  \centering
  \begin{tabular}{lll}
    \toprule
    \cmidrule(r){1-2}
    Layer (type)  & Output Shape & Param \# \\
    \midrule
   lstm\_1 (LSTM)& (None, 1000) & 4,212,000       \\
    dense\_1 (Dense) &  (None, 1) & 1,001 \\    
    \midrule
    Total params: 4,213,001\\
    Trainable params: 4,213,001\\
    Non-trainable params: 0\\
    \bottomrule

  \end{tabular}
  \label{tab:table}
\end{table}

For the  \texttt{model using options in order}, the MSE of the predicted value vs the actual value is $5.632$. For the \texttt{model using every third option}, the MSE is $7.797$. The model training loss and the predictions vs the actual VIX values are shown below. Both models give us an acceptable performance.

\begin{figure}[h]
\begin{subfigure}{0.5\textwidth}
\includegraphics[width=0.9\linewidth, height=5cm]{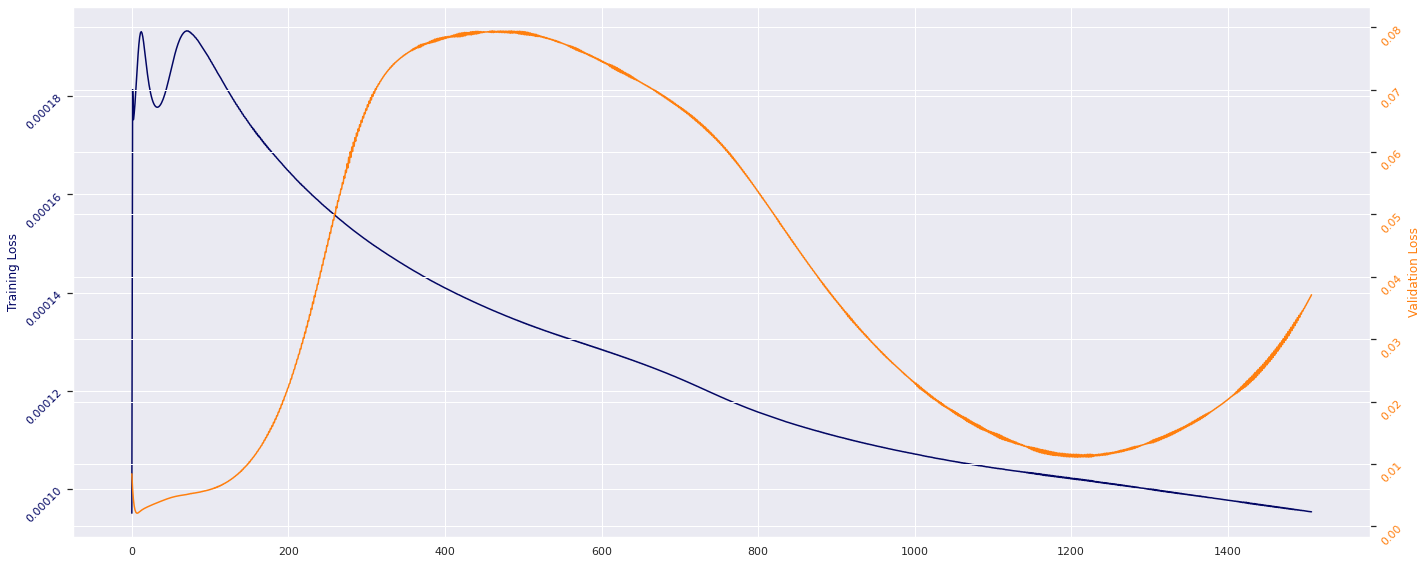}
\caption{Options in Order: Model Training Loss}
\label{fig:subim1}
\end{subfigure}
\begin{subfigure}{0.5\textwidth}
\includegraphics[width=0.9\linewidth, height=5cm]{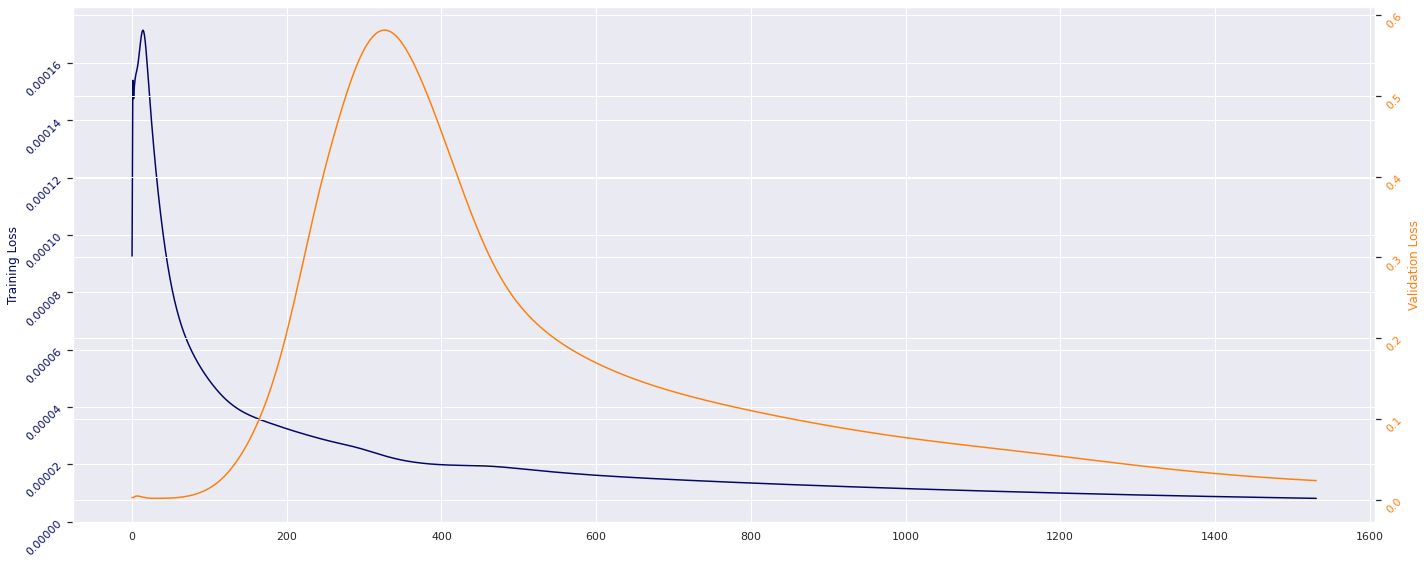}
\caption{Every Third Option: Model Training Loss}
\label{fig:subim3}
\end{subfigure}
\end{figure}    

\begin{figure}[h]
\begin{subfigure}{0.5\textwidth}
\includegraphics[width=0.9\linewidth,height=5cm]{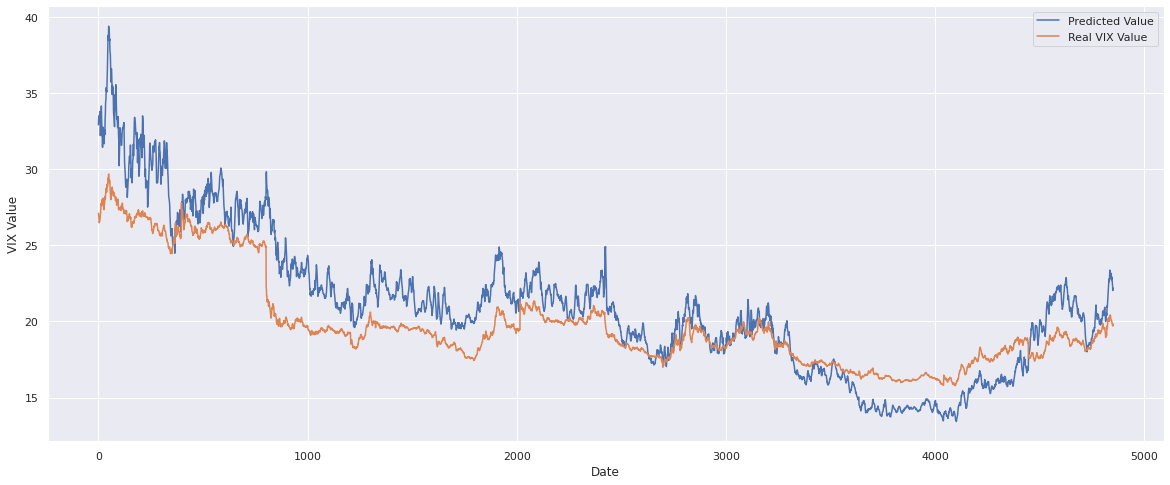}
\caption{Options in Order: Predictions vs Real VIX by Quote\_time}
\label{fig:subim1}
\end{subfigure}
\begin{subfigure}{0.5\textwidth}
\includegraphics[width=0.9\linewidth,height=5cm]{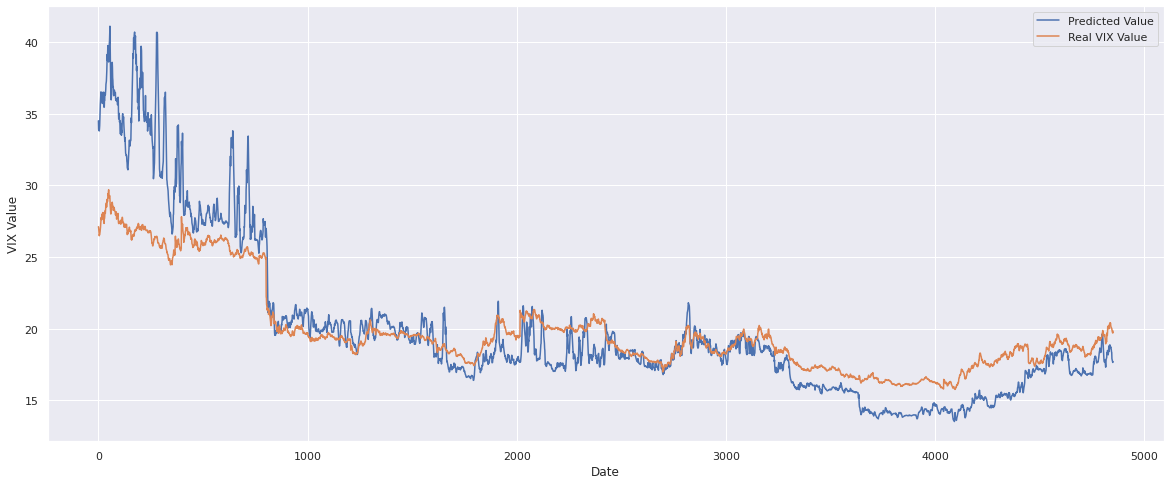}
\caption{Every Third Option: Predictions vs Real VIX by Quote\_time}
\label{fig:subim4}
\end{subfigure}
\end{figure}    

\begin{figure}[h]
\begin{subfigure}{0.5\textwidth}
\includegraphics[width=0.9\linewidth, height=5cm]{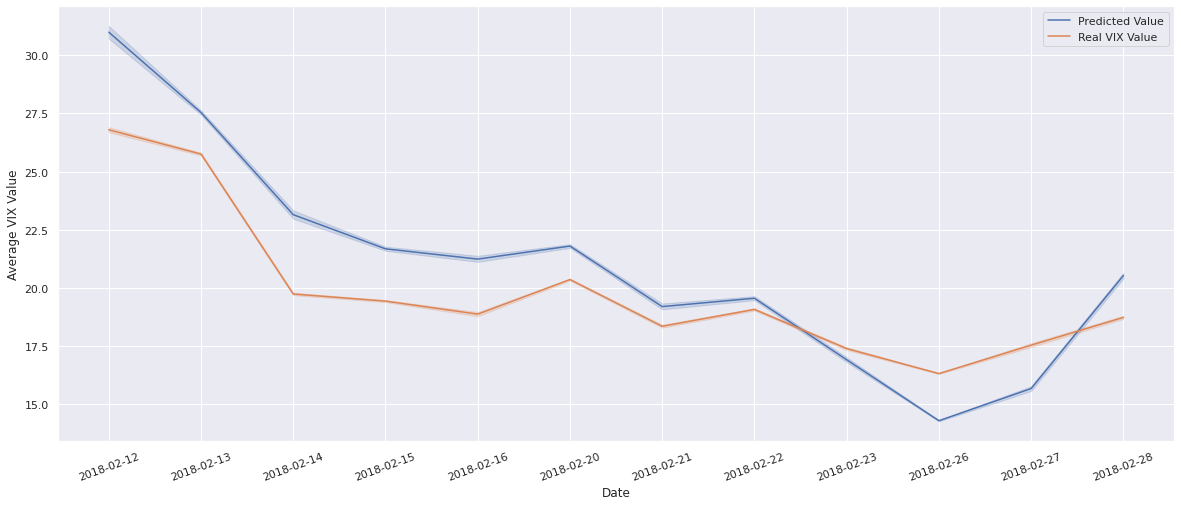}
\caption{Options in Order: Predictions vs Real VIX by Date}
\label{fig:subim1}
\end{subfigure}
\begin{subfigure}{0.5\textwidth}
\includegraphics[width=0.9\linewidth, height=5cm]{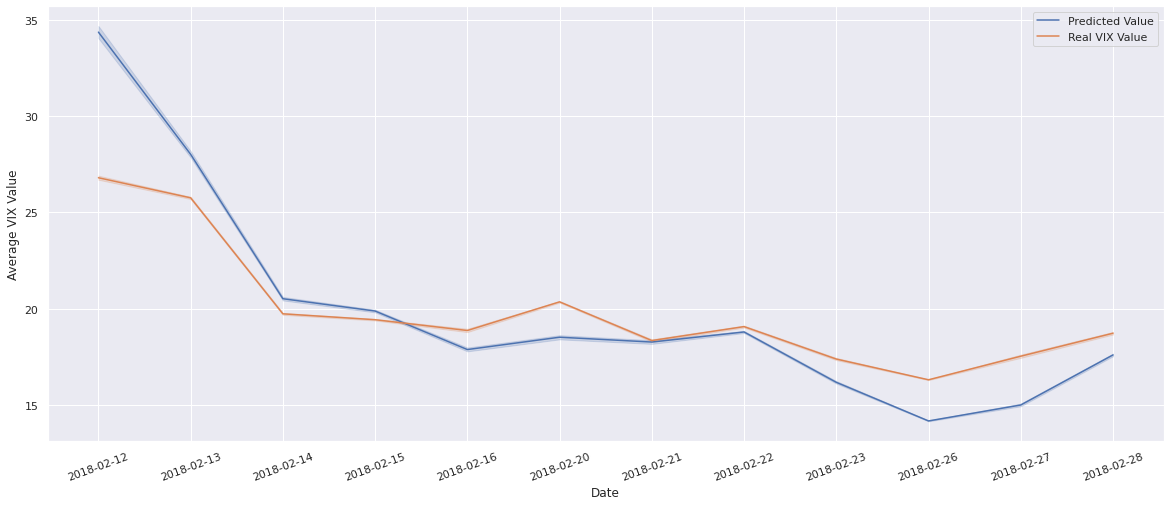}
\caption{Every Third Option: Predictions vs Real VIX by Date}
\label{fig:subim5}
\end{subfigure}

\end{figure}
    
\subsubsection {Model 2: The Multi-Layer Model} 
The multi-layer model contains two LSTM layers and one dropout layer. The LSTM layers have 500 output units, and the $30\%$ dropout layers are for the purpose of overfitting prevention. The model also contains another LSTM layer with one output unit, and a dense layer to return our prediction value (see Table 3). 
  
\begin{table}
 \caption{LSTM Model: Multi-Layer}
  \centering
  \begin{tabular}{lll}
    \toprule
    \cmidrule(r){1-2}
    Layer (type)  & Output Shape & Param \# \\
    \midrule
   lstm\_1 (LSTM)& (None, 10, 500) & 1,106,000       \\
    dropout\_1 (Dropout) &  (None, 10, 500) & 0 \\
   lstm\_2 (LSTM)& (None, 10, 500) & 2,002,000       \\
    dropout\_2 (Dropout) &  (None, 10, 500) & 0 \\  
   lstm\_3 (LSTM)& (None, 500) & 2,002,000       \\
    dense (Dense) &  (None, 1) & 501 \\    
    \midrule
    Total params: 5,110,501\\
    Trainable params: 5,110,501\\
    Non-trainable params: 0\\
    \bottomrule

  \end{tabular}
  \label{tab:table}
\end{table}
The \texttt{model using options in order} is the only model that has not triggered early stopping at epoch $1500$. Its MSE of the predicted value against the actual value is $17.918$, the worst performance among all. For \texttt{model using every third option}, the MSE of the predicted value against the actual value is $2.963$. The model training loss and the predictions vs the actual VIX are shown below. While the performance of model using options in order is not as good, the model using Every Third Option give us the best performance among the four.
\begin{figure}[h]
\begin{subfigure}{0.5\textwidth}
\includegraphics[width=0.9\linewidth, height=5cm]{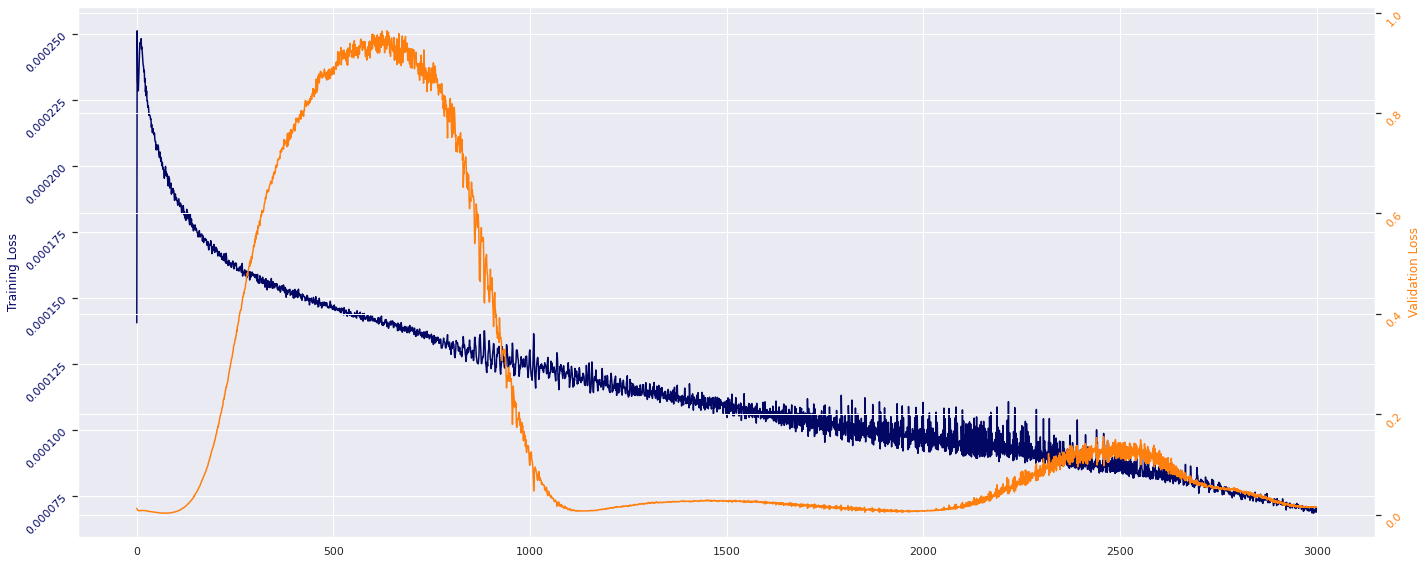}
\caption{Options in Order: Model Training Loss}
\label{fig:subim1}
\end{subfigure}
\begin{subfigure}{0.5\textwidth}
\includegraphics[width=0.9\linewidth, height=5cm]{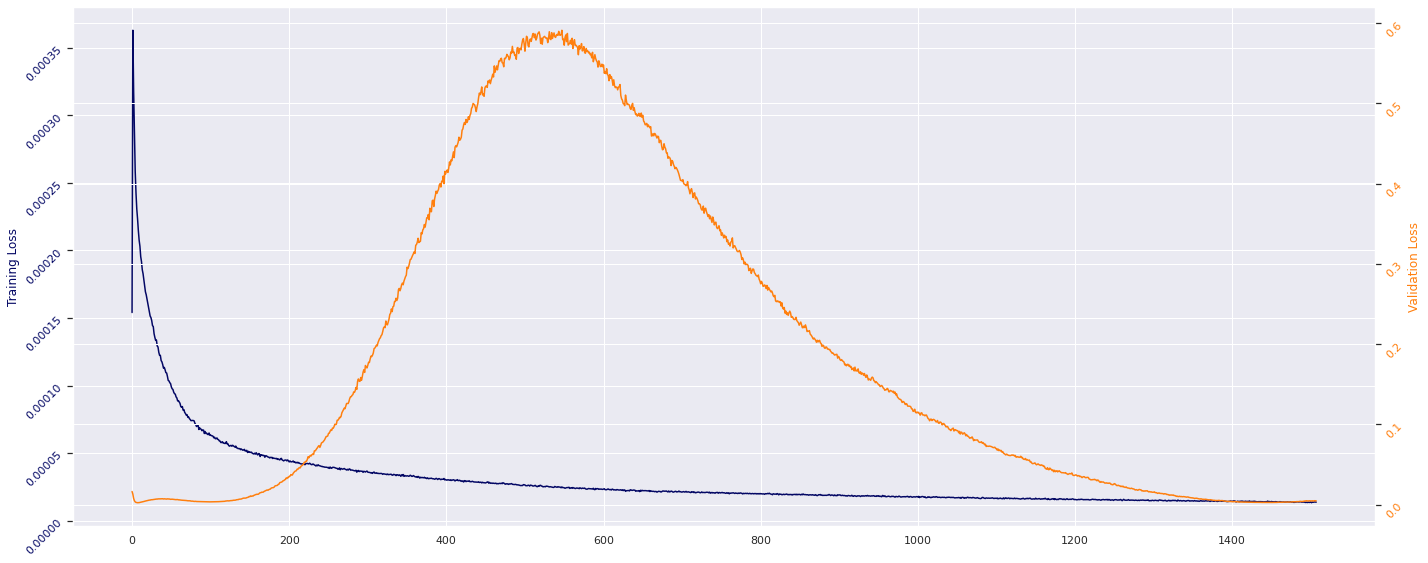}
\caption{Every Third Option: Model Training Loss}
\label{fig:subim6}
\end{subfigure}
\end{figure}    
\begin{figure}[h]
\begin{subfigure}{0.5\textwidth}
\includegraphics[width=0.9\linewidth, height=5cm]{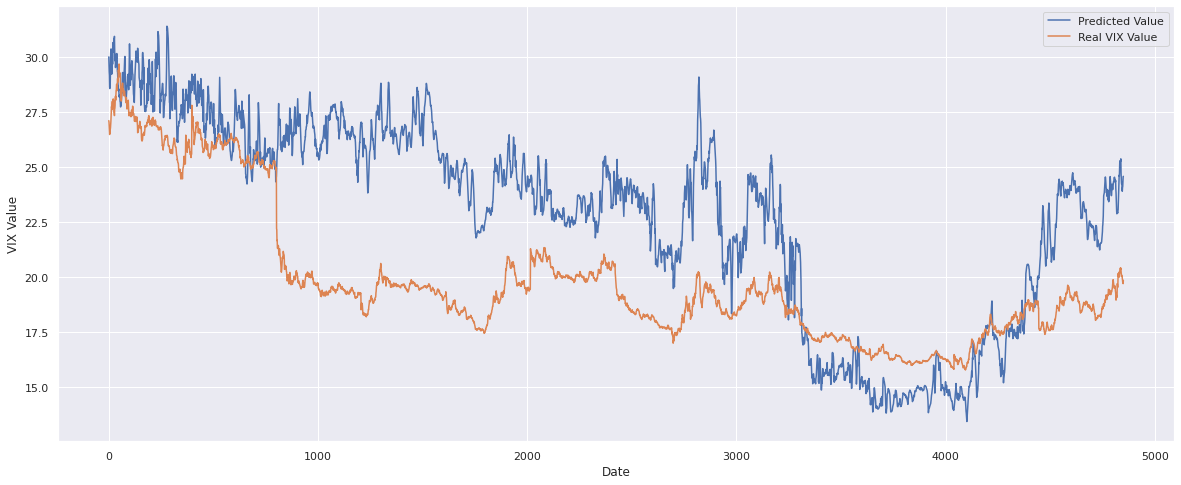}
\caption{Options in Order: Predictions vs Real VIX by Quote\_time}
\label{fig:subim1}
\end{subfigure}
\begin{subfigure}{0.5\textwidth}
\includegraphics[width=0.9\linewidth, height=5cm]{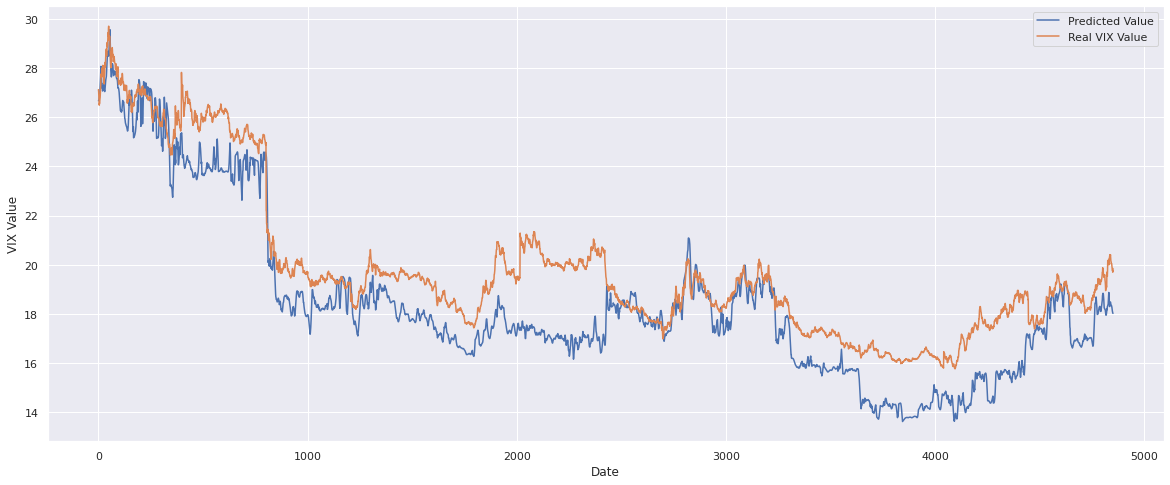}
\caption{Every Third Option: Predictions vs Real VIX by Quote\_time}
\label{fig:subim7}
\end{subfigure}
\end{figure}    

\begin{figure}[h]
\begin{subfigure}{0.5\textwidth}
\includegraphics[width=0.9\linewidth, height=5cm]{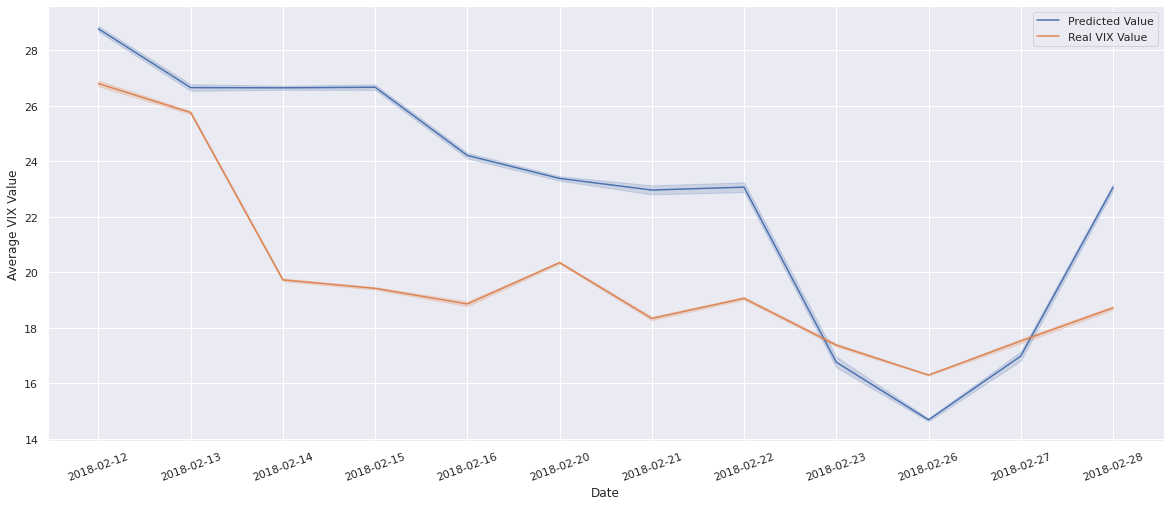}
\caption{Options in Order: Predictions vs Real VIX by Date}
\label{fig:subim1}
\end{subfigure}
\begin{subfigure}{0.5\textwidth}
\includegraphics[width=0.9\linewidth, height=5cm]{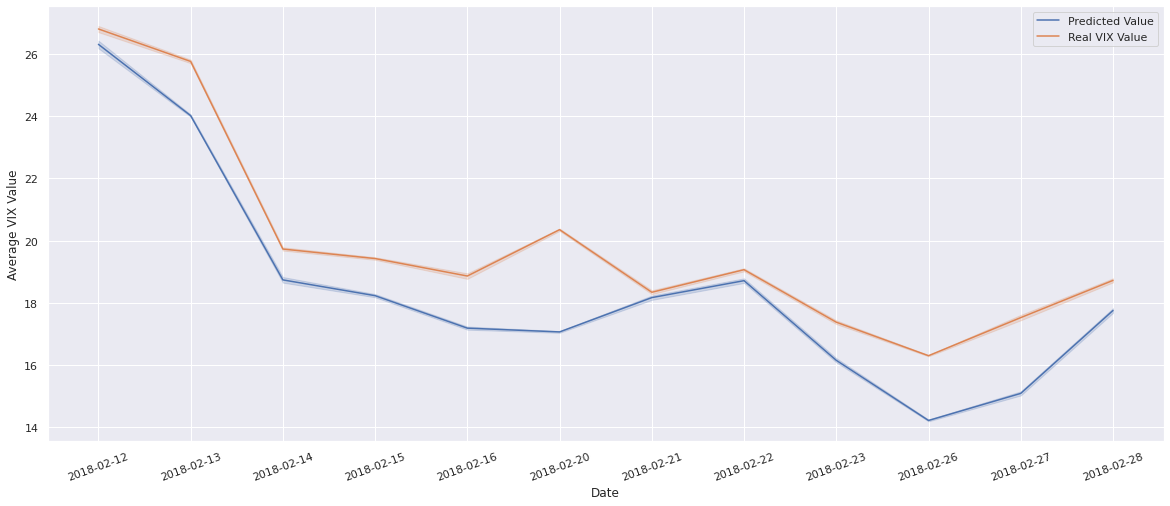}
\caption{Every Third Option: Predictions vs Real VIX by Date}
\label{fig:subim8}
\end{subfigure}
\end{figure}

\section{VIX futures}
Monthly VIX futures are futures contracts with monthly expirations, being cash-settled with the settlement amount being determined on the Wednesday that is thirty days prior to the third Friday of the calendar month immediately following the month in which the contract expires. The final settlement value is determined in a Special Opening Auction with the VIX normally being calculated based on actual traded option prices.

The fair value of a VIX futures contract is the square root of the implied variance minus an adjustment representing the concavity of the forward position (Hancock(2012)).

There has been strong evidence observed in the market that VIX and SPX option prices are negatively correlated with each other, meaning that including VIX in a portfolio will be a good hedging choice. The VIX exposure for downside protection is illustrated by constructing two hypothetical portfolios, one holding only the SPY and the other holding 90 percent of the wealth in SPY and 10 percent in the VIX. In the study by Tim Leung, when negative news stroke the market, like credit rating downgrades by Standard and Poor's in 2011, the former portfolio would lose up to 10 percent with a much more volatile trajectory while the latter one (VIX included) would have yield a positive return, remaining stable throughout the period when markets collapsed. But the fact that VIX is not a traded asset, but rather an anticipated benchmark for the US stock market volatility leads to the existence of VIX derivatives like VIX futures and VIX options. Most of the time volatility exposure is achieved using these two products. These instruments give speculators to establish a position on market volatility with a long position in the VIX futures contract meaning a bet on the increase of market volatility and a decline in the market index.

\section{Relationship between VIX and VIX futures}
Numerous research has been done on the VIX and VIX futures and it is obvious that these two are highly correlated since the settlement price of VIX futures is the future variance adjusted downward by the concavity. Regressing the returns on VIX futures against returns on VIX yields a slope coefficient statistically significant less than 1 and a negative intercept for all contracts with different maturities, showing that futures returns are less volatile than spot returns and that when the spot price does not move, futures prices still tend to fall as futures prices will converge to the spot price as the maturity date approaches. 

\section{LSTM for predicting VIX futures}
\subsection{Data and Methodology} 
The input data is the same as used earlier. For the output data, we substitute the VIX index with VIX futures values starting from 2018-01-02 to 2018-02-28. Specifically for VIX futures, we utilize minute-by-minute data from 9:30:00 am to 16:14:00 p.m. every day in order to match the time-stamps of the S\&P500 option prices data. Again, we choose 52 out-of-the-money options (every third option as per strike price), including both near-term and next term, as features.  

\subsection{Model 1: The Flat Model} 
The first model is the same as considered before, having an LSTM layer with 1000 output units, and a dense layer to output the predicted value (see Table 4). 
\begin{table}
 \caption{LSTM Model 1}
  \centering
  \begin{tabular}{lll}
    \toprule
    \cmidrule(r){1-2}
    Layer (type)  & Output Shape & Param \# \\
    \midrule
   lstm\_1 (LSTM)& (None, 1000) & 4,212,000       \\
    dense\_1 (Dense) &  (None, 1) & 1,001 \\    
    \midrule
    Total params: 4,213,001\\
    Trainable params: 4,213,001\\
    Non-trainable params: 0\\
    \bottomrule

  \end{tabular}
  \label{tab:table}
\end{table}
For \texttt{model using every option in order}, the MSE of the predicted value against the actual value is $22.295$ and the MAE is 4.417. For \texttt{model using every third option}, the MSE of the predicted value against the actual value is $19.051$ and MAE is $4.070$. We can see that the MSEs for both models are significantly larger than previous performances when predicting VIX values and both triggered early stopping at epoch $1757$ and epoch $1500$ respectively.
    \begin{figure}[h]
    \begin{subfigure}{0.5\textwidth}
    \includegraphics[width=0.9\linewidth, height=5cm]{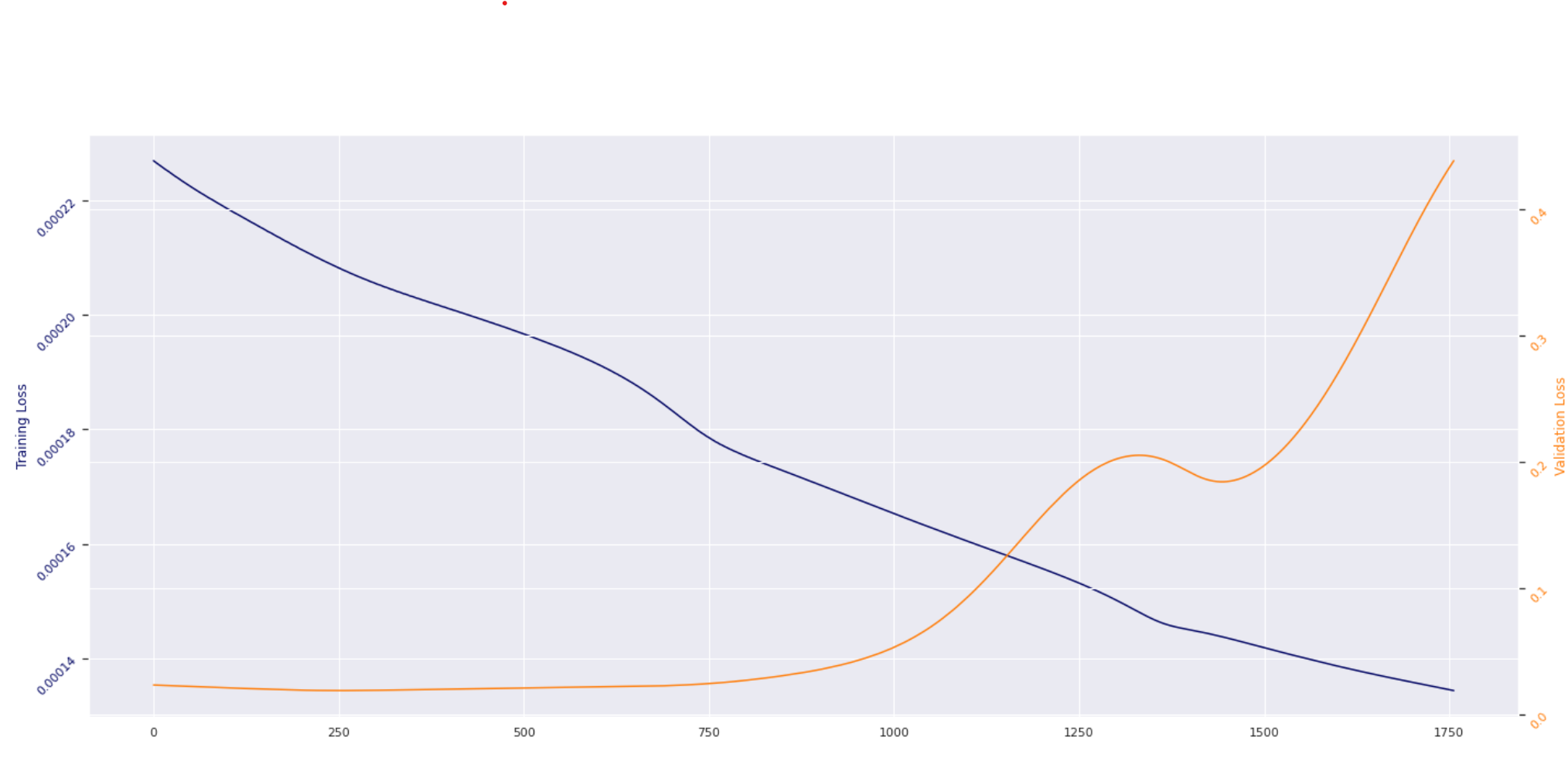}
    \caption{Every Option in Order: Model Training Loss}
    \label{fig:subim1}
    \end{subfigure}
    \begin{subfigure}{0.5\textwidth}
    \includegraphics[width=0.9\linewidth, height=5cm]{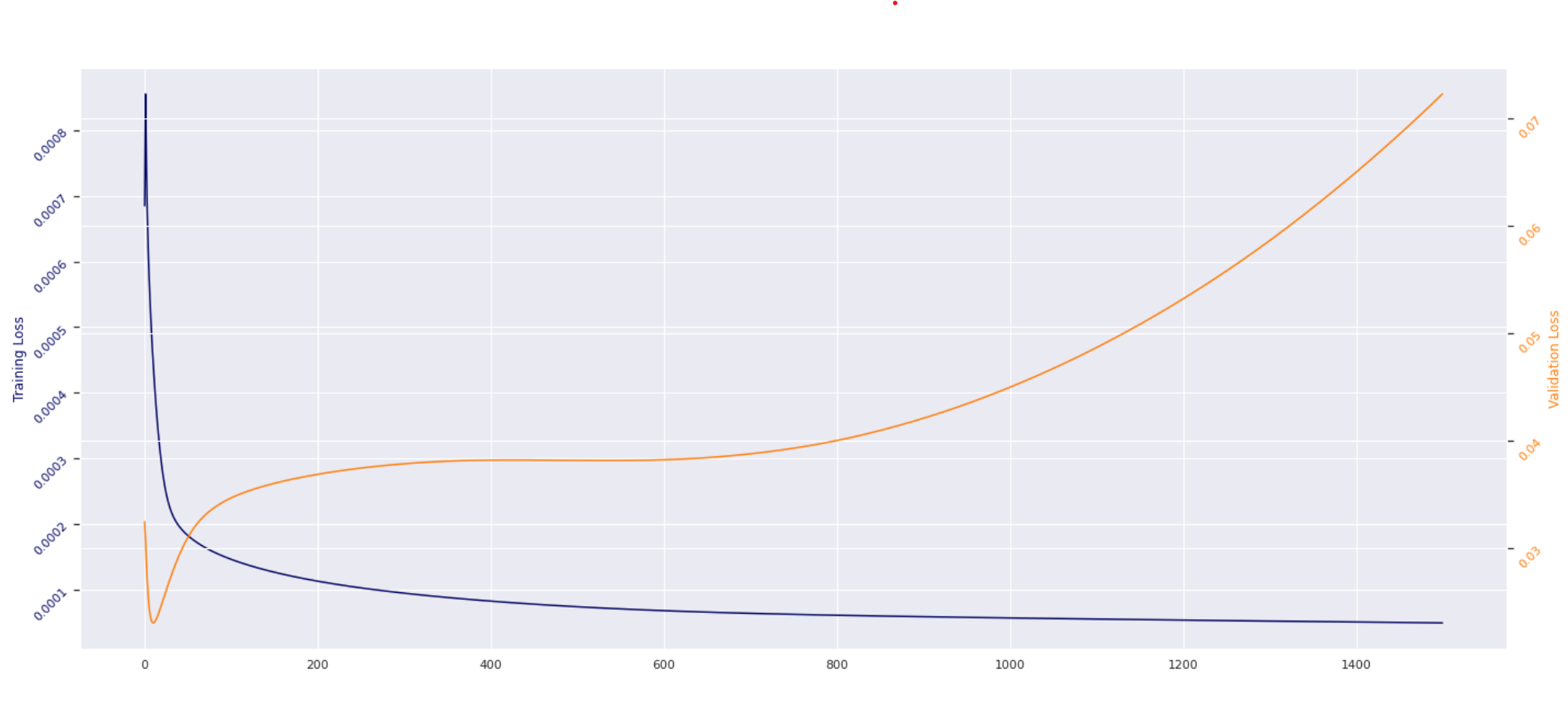}
    \caption{Every Third Option: Model Training Loss}
    \label{fig:subim9}
    \end{subfigure}
    \end{figure}    
    \begin{figure}[h]
    \begin{subfigure}{0.5\textwidth}
    \includegraphics[width=0.9\linewidth, height=5cm]{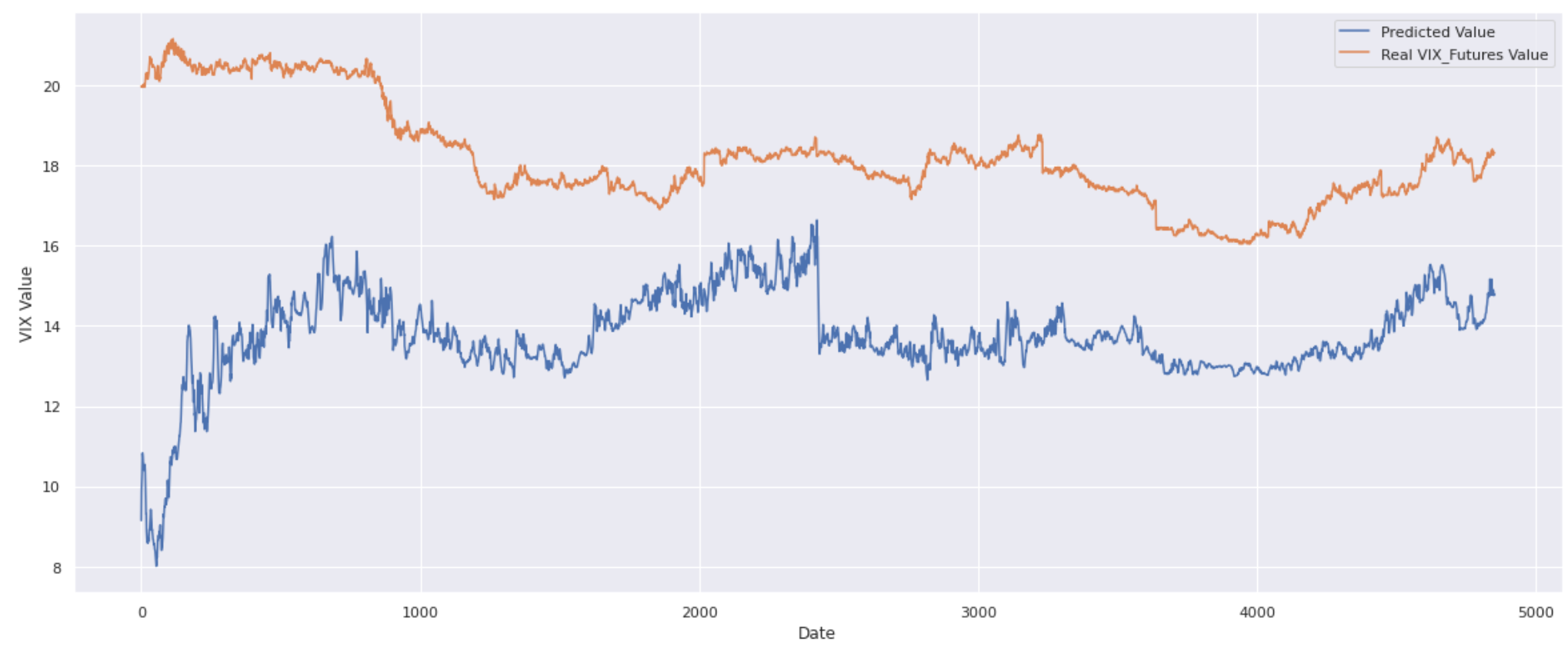}
    \caption{Every Option in Order: Predictions vs actual VIX}
    \label{fig:subim1}
    \end{subfigure}
    \begin{subfigure}{0.5\textwidth}
    \includegraphics[width=0.9\linewidth, height=5cm]{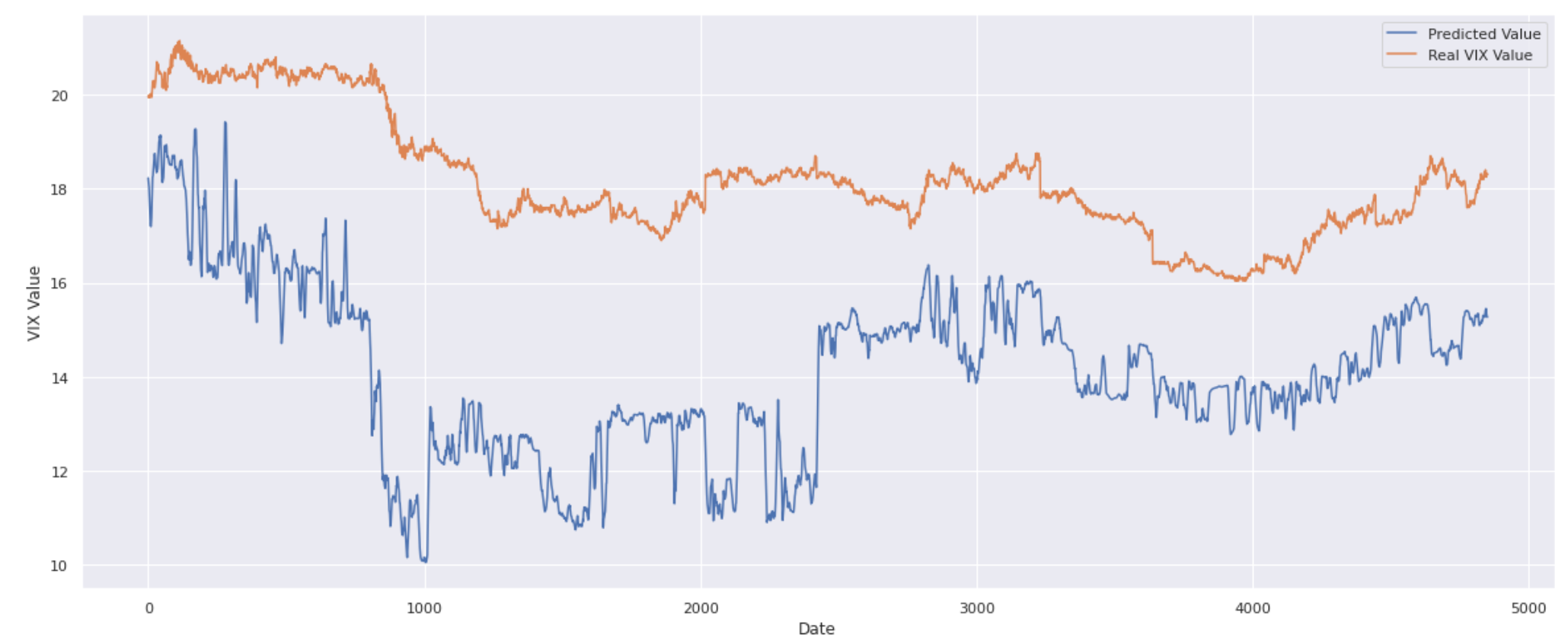}
    \caption{Every Third Option: Predictions vs actual VIX}
    \label{fig:subim10}
    \end{subfigure}
    \end{figure}    
    
    \begin{figure}[h]
    \begin{subfigure}{0.5\textwidth}
    \includegraphics[width=0.9\linewidth, height=5cm]{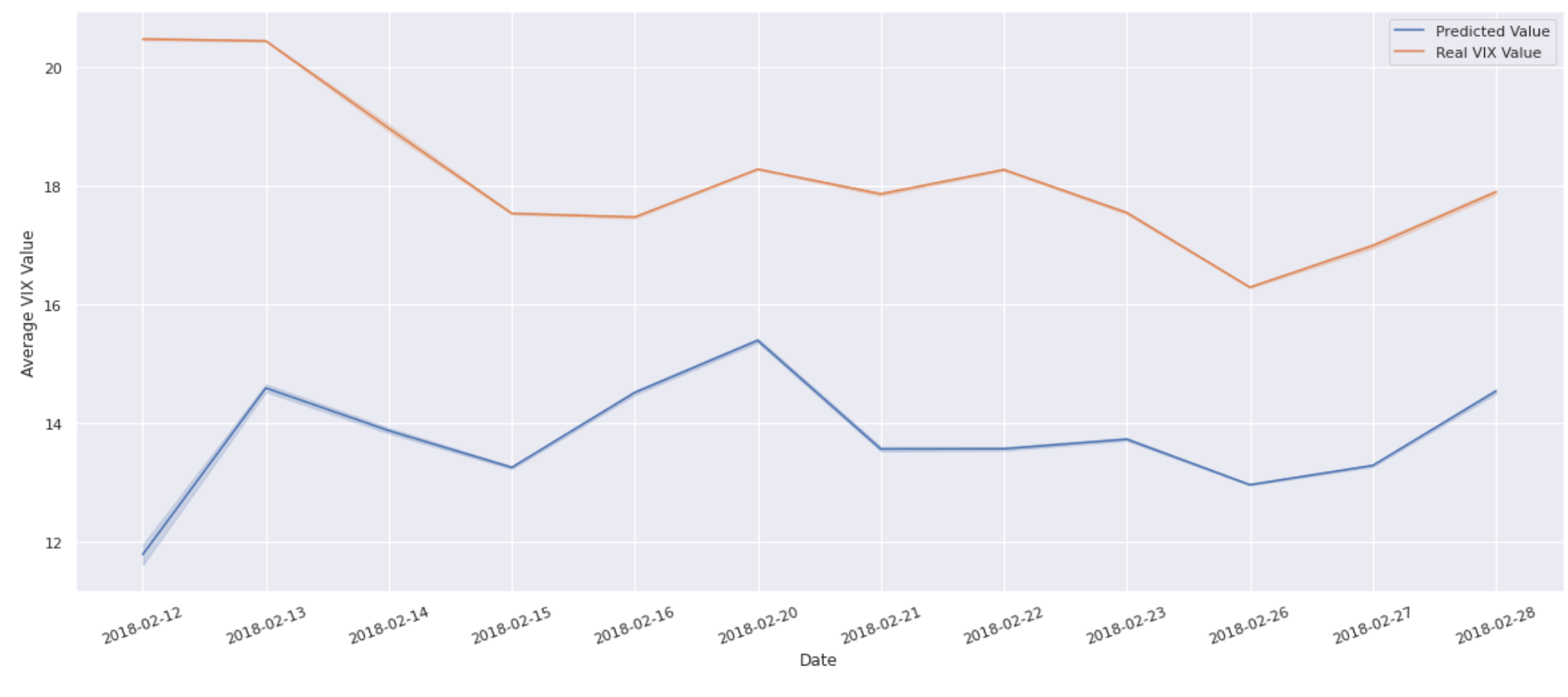}
    \caption{Every Option in Order: Predictions vs actual VIX}
    \label{fig:subim1}
    \end{subfigure}
    \begin{subfigure}{0.5\textwidth}
    \includegraphics[width=0.9\linewidth, height=5cm]{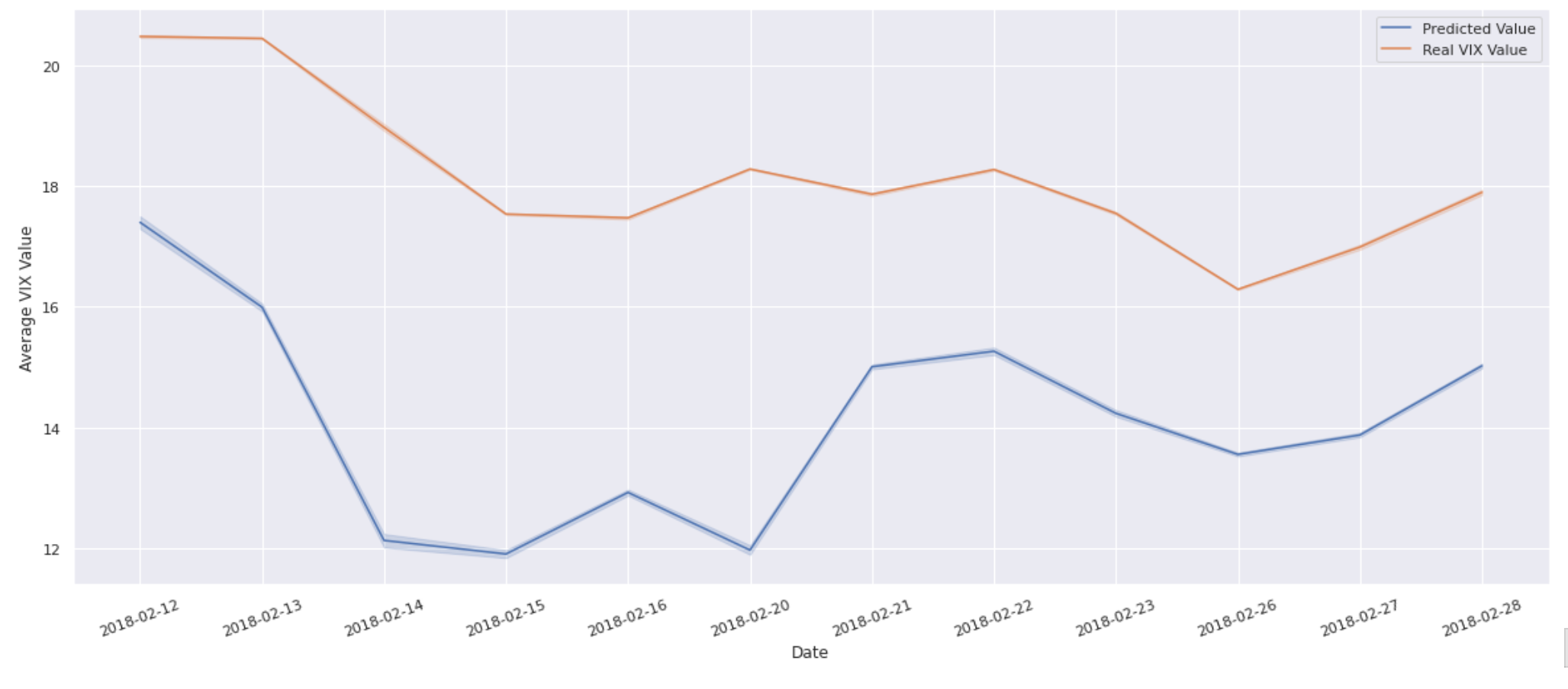}
    \caption{Every Third Option: Predictions vs actual VIX}
    \label{fig:subim11}
    \end{subfigure}
    \end{figure}

\subsection{Model 2: The Multi-Layer Model} 
We use the same architecture for the neural network as before (see Table 3). The \texttt{model using every option in order} triggered early stopping at epoch $1500$. Its MSE of the predicted value against the actual value is $4.437$ and MAE being $1.724$, achieving the best performance among all. For \texttt{model using every third option}, it also triggered early stopping at epoch $1500$. The MSE of the predicted value against the actual value is $28.896$ and MAE being $5.014$, achieving the worst performance among all models. The model using every options gives us the best performance among the four considered versions.
    \begin{figure}[h]
    \begin{subfigure}{0.5\textwidth}
    \includegraphics[width=0.9\linewidth, height=5cm]{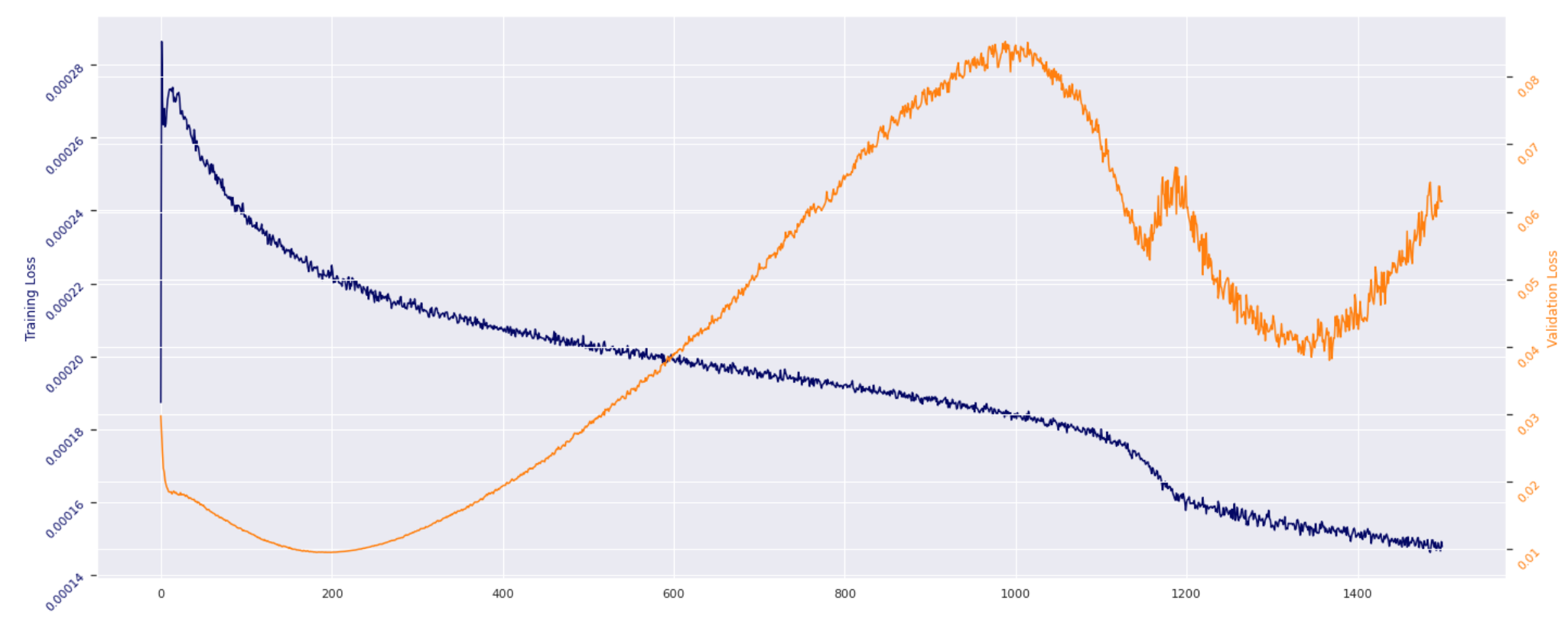}
    \caption{Every Option in Order: Model Training Loss}
    \label{fig:subim1}
    \end{subfigure}
    \begin{subfigure}{0.5\textwidth}
    \includegraphics[width=0.9\linewidth, height=5cm]{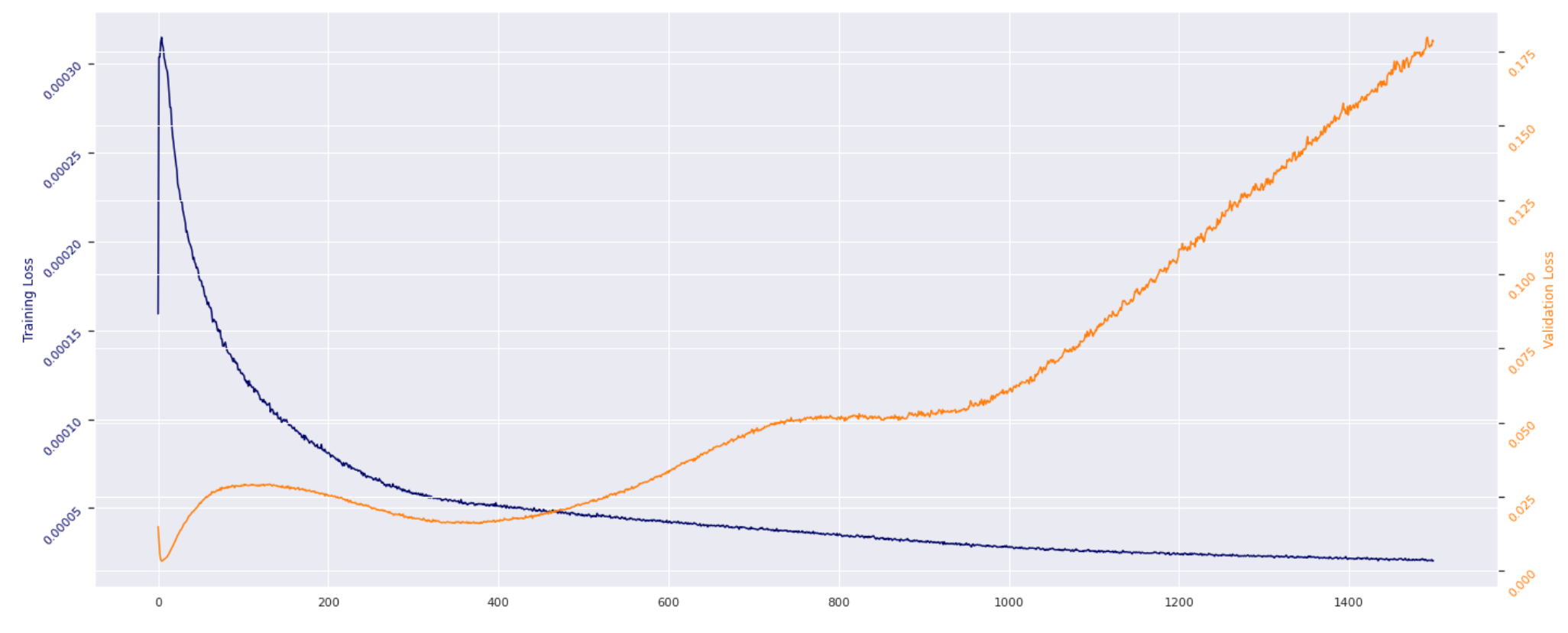}
    \caption{Every Third Option: Model Training Loss}
    \label{fig:subim12}
    \end{subfigure}
    \end{figure}    
    \begin{figure}[h]
    \begin{subfigure}{0.5\textwidth}
    \includegraphics[width=0.9\linewidth, height=5cm]{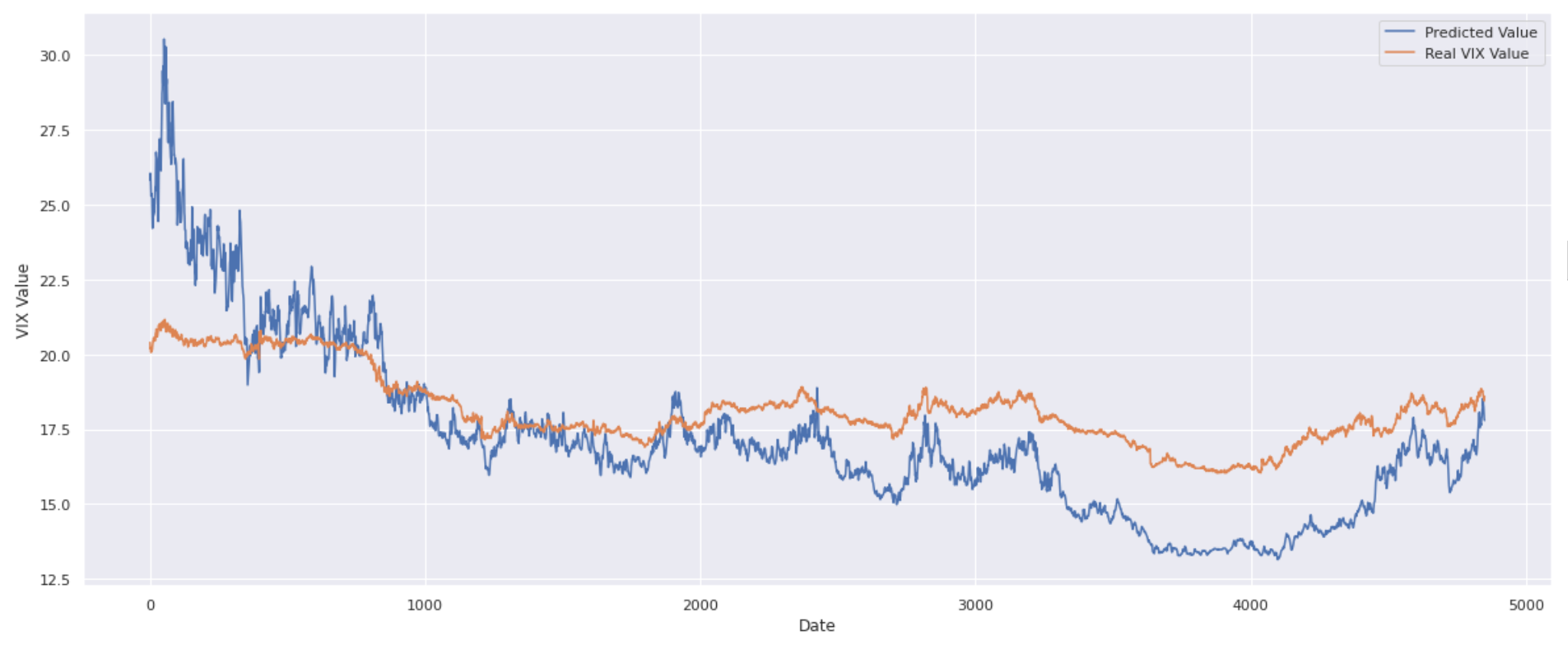}
    \caption{Every Option in Order: Predictions vs actual VIX}
    \label{fig:subim1}
    \end{subfigure}
    \begin{subfigure}{0.5\textwidth}
    \includegraphics[width=0.9\linewidth, height=5cm]{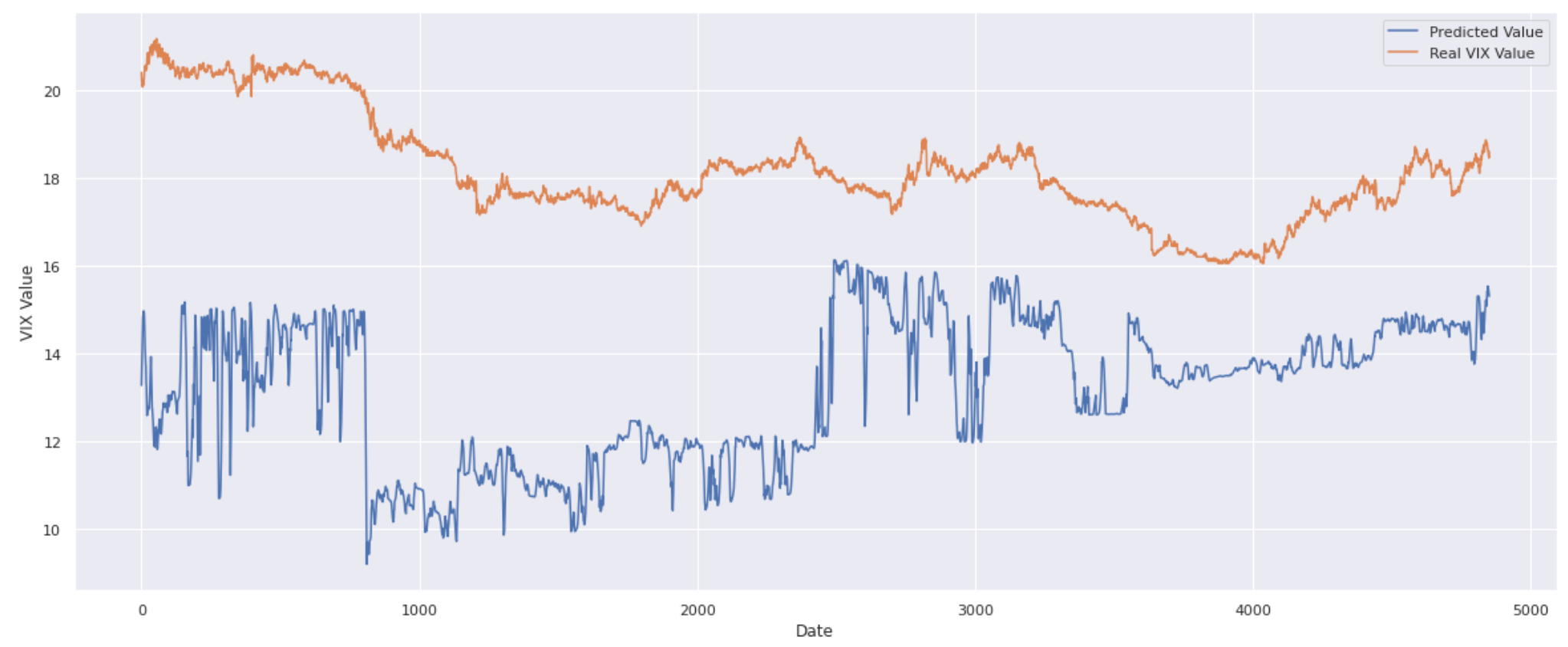}
    \caption{Every Third Option: Predictions vs Real VIX by Quote\_time}
    \label{fig:subim13}
    \end{subfigure}
    \end{figure}    
    
    \begin{figure}[h]
    \begin{subfigure}{0.5\textwidth}
    \includegraphics[width=0.9\linewidth, height=5cm]{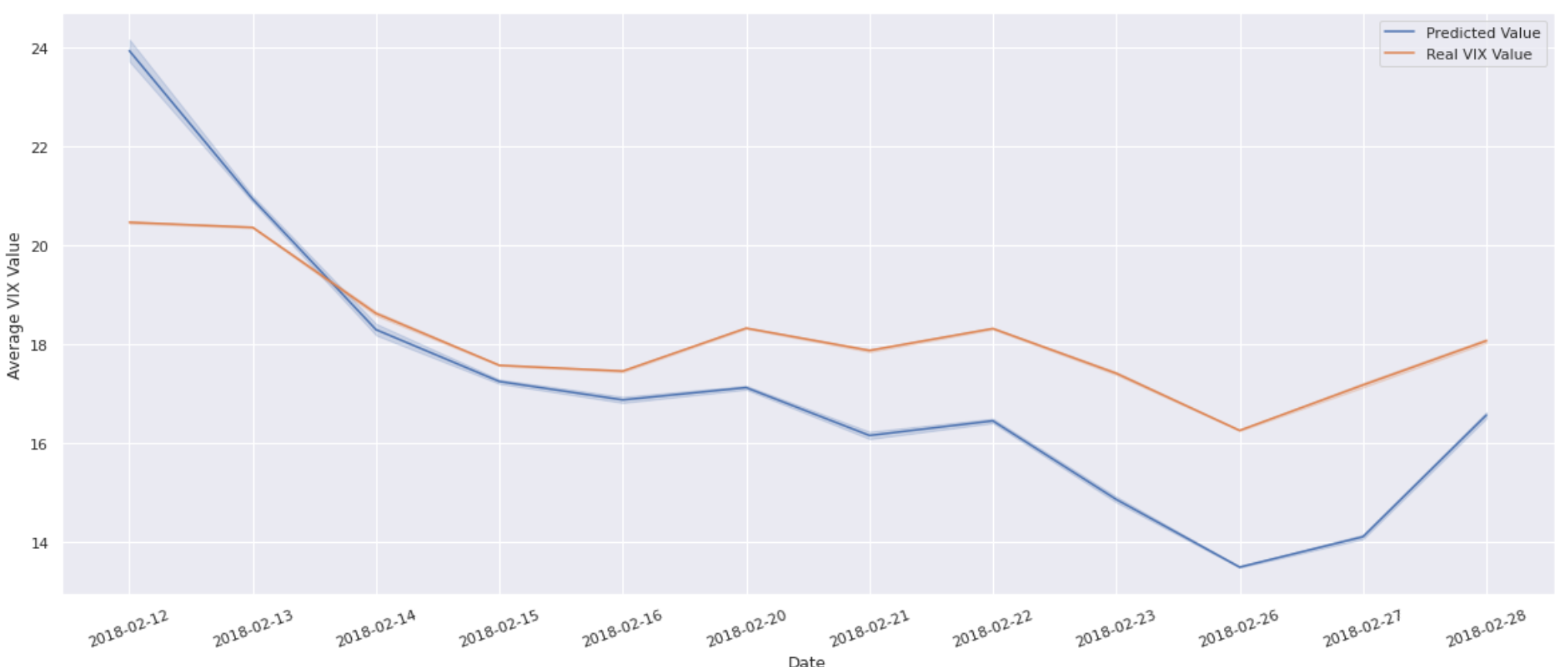}
    \caption{Every Option in Order: Predictions vs actual VIX}
    \label{fig:subim1}
    \end{subfigure}
    \begin{subfigure}{0.5\textwidth}
    \includegraphics[width=0.9\linewidth, height=5cm]{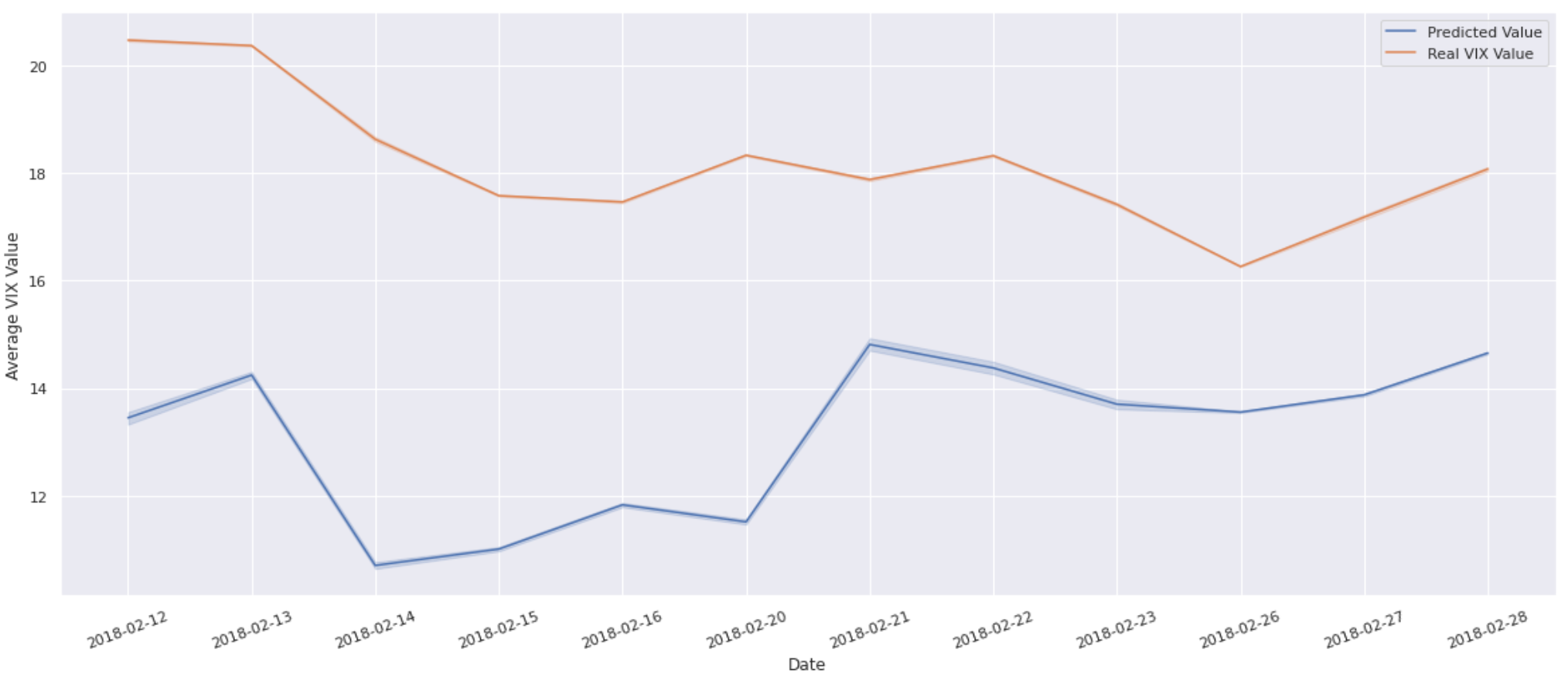}
    \caption{Every Third Option: Predictions vs actual VIX}
    \label{fig:subim14}
    \end{subfigure}
    \end{figure}

\section{Conclusion and Summary}
\subsection{VIX replication and prediction} 
In this study, we first performed a VIX replication using both daily and intraday minute-by-minute data. After selecting options according to the official method, we selected subsets of them through our subset selection methods. We replicated the VIX using the official formulas and analysed the results. We then selected the subset with the best performance as the input of our feed-forward neural network and LSTM models.

The performance of the basic neural network model is acceptable. After adding more features, the model gives an even better performance. For LSTM models, we tried two different model architectures each with two different input options. Among those, the flat model using the options in order as well as the multiple-layer model using every-third options have better performance.

Thus, our work shows that instead of using all options selected by the CBOE's methodology, which are about 300 options per minute, we can apply machine learning and deep learning techniques to replicate the VIX index by using a smaller subset containing only 52 options per minute (including both near-term and next-term options) with a small MSE. 

For further studies, one can develop a selection methodology that is not dependent on the CBOE's methodology in the first place, but that uses a neural network for choosing an optimal set of otpions.
\subsection{VIX futures replication and prediction} 
Using the same network architecture as before, we are predicting VIX futures. Instead of using all options selected by CBOE's methodology and adjusting it with a downward concavity term to obtain the VIX futures index, our work shows that using deep learning models based on a LSTM architecture to predict the future index with only a small subset of 52 option prices per minute is feasible. The multi layer LSTM model \texttt{using every option price in order} yielded rather satisfatory results (MSE $4.437$), but the other three models have performances not quite satisfactory, so we conclude that a multi-layer LSTM model works better in predicting the results than a flat LSTM model. Also, it is of importance that we choose the proper input data, consecutive option prices, as opposed to every third option prices, turns out better in the prediction of VIX futures.

\newpage
\bibliographystyle{unsrt}  


\appendix
\section{Appendix}

\begin{figure}[h]
\includegraphics[width=0.32\linewidth, height=5.5cm]{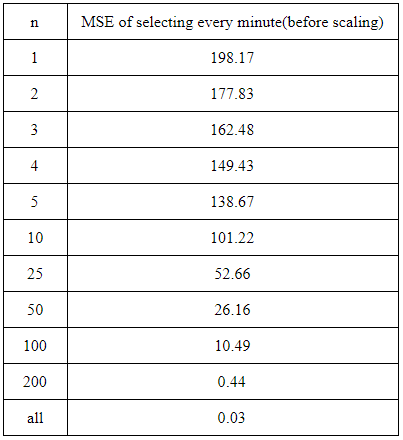}
\centering
\caption{MSE of selecting a subset of options every minute with different numbers of options (n) to replicate VIX (before scaling)}
\label{fig:subim1}
\end{figure}
\begin{figure}[h]
\begin{center}
\includegraphics[width=0.8\linewidth, height=7cm]{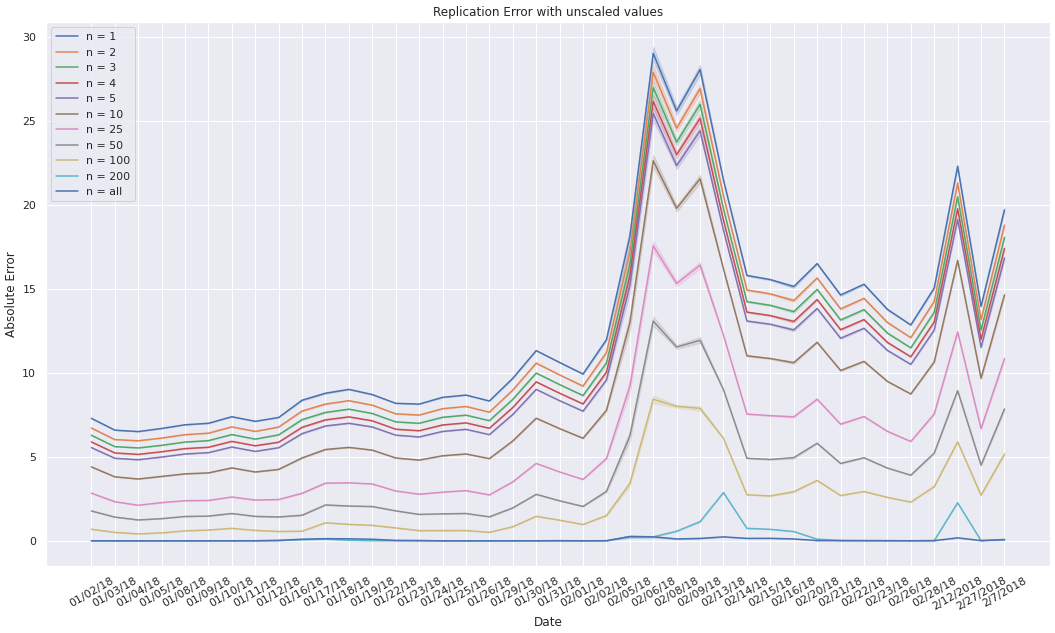}
\caption{Error of selecting a subset of options every minute with different numbers of options (n) to replicate VIX (before scaling)}
\label{fig:subim1a}
\end{center}
\end{figure}

\begin{figure}[h]
\includegraphics[width=0.34\linewidth, height=5.3cm]{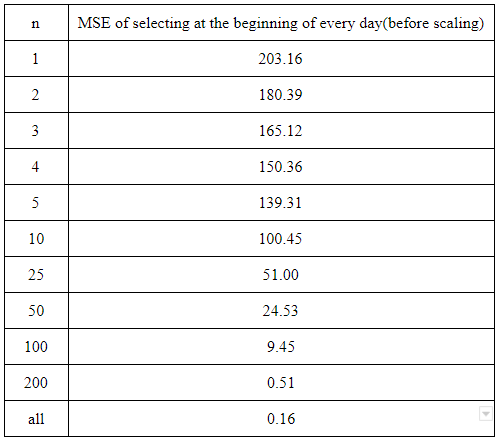}
\centering
\caption{MSE of selecting a subset of options at the beginning of every day with different numbers of options (n) to replicate VIX (before scaling)}
\label{fig:subim1b}
\end{figure}
\begin{figure}[h]
\begin{center}
\includegraphics[width=0.8\linewidth, height=7cm]{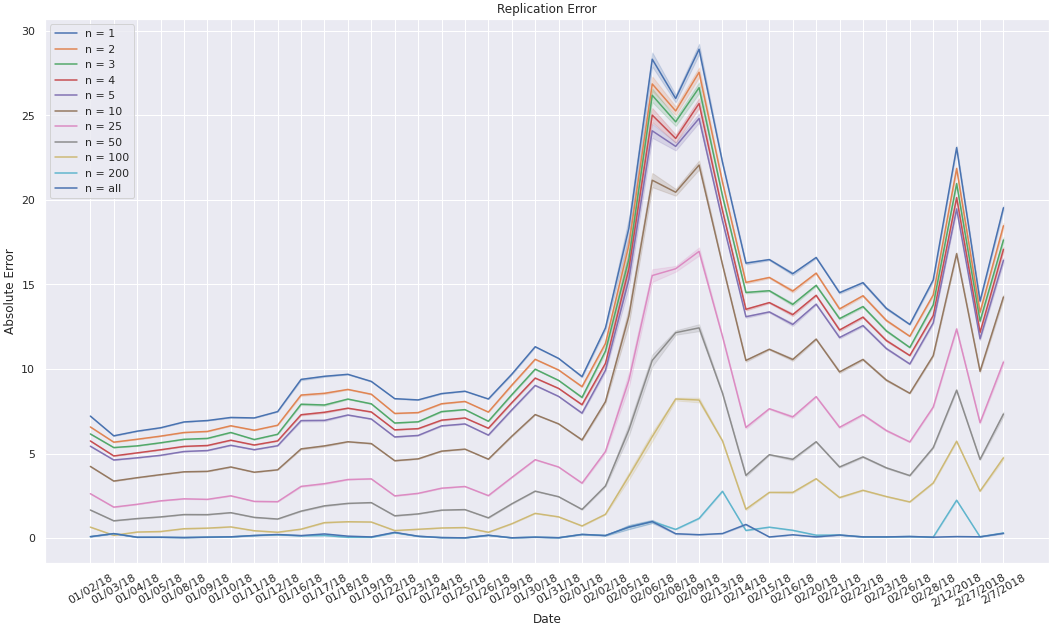}
\caption{Error of selecting a subset of options at the beginning of every day with different numbers of options(n) to replicate VIX (before scaling)}
\label{fig:subim1c}
\end{center}
\end{figure}

\end{document}